\documentclass[acmsmall]{acmart}

\AtBeginDocument{%
  \providecommand\BibTeX{{%
    Bib\TeX}}}

\setcopyright{cc}
\setcctype{by-sa}
\acmDOI{10.1145/3729379}
\acmYear{2025}
\acmJournal{PACMSE}
\acmVolume{2}
\acmNumber{FSE}
\acmArticle{FSE109}
\acmMonth{7}
\received{2024-09-13}
\received[accepted]{2025-04-01}

\usepackage{natbib}

\usepackage[frozencache,cachedir=.]{minted}

\usepackage[noend]{algpseudocode}
\usepackage{graphicx}
\usepackage[tikz]{bclogo}
\usepackage{wrapfig}
\usepackage{textcomp}
\usepackage{xcolor}
\usepackage{arydshln}
\usepackage{enumitem}
\usepackage{soul}
\usepackage{cleveref}
\usepackage[belowskip=0pt,aboveskip=5pt]{subcaption}
\usepackage{listings}
\usepackage{xspace}
\usepackage[most]{tcolorbox}
\usepackage{multirow}
\usepackage{color}
\usepackage{url}
\usepackage[export]{adjustbox}
\usepackage[linesnumbered,ruled,vlined]{algorithm2e}
\usepackage{amsmath,amsfonts}
\definecolor{bubblegum}{rgb}{0.99, 0.76, 0.8}
\definecolor{cambridgeblue}{rgb}{0.64, 0.76, 0.68}
\newcommand{\approach}{\textsc{AlphaTrans}\xspace}

\newcommand{\llm}{DeepSeek-Coder-33b-Instruct\xspace}
\newcommand{\gptfo}{GPT-4o\xspace}
\definecolor{ibmblue}{RGB}{63,97,246}
\newcommand{\rebuttal}[1]{\textcolor{black}{#1}}

\newcommand{\Space}[1]{}

\setmintedinline{breaklines}
\newcommand{\java}[1]{{\small \mintinline[breakanywhere]{java}{#1}}}

\SetKwInput{KwInputs}{Inputs}                % Set the Input
\SetKwInput{KwOutputs}{Outputs}              % set the Output
\SetKwInput{KwOutput}{Output}              % set the Output

\def\BibTeX{{\rm B\kern-.05em{\sc i\kern-.025em b}\kern-.08em
    T\kern-.1667em\lower.7ex\hbox{E}\kern-.125emX}}

\newcommand{\mybox}[1]{\begin{tcolorbox}[enhanced, frame hidden, boxsep=0pt]\emph{#1}\end{tcolorbox}}

\newcommand*\circled[2][fill=black]{\tikz[baseline=(char.base)]{
    \footnotesize
    \node[shape=circle, #1, inner sep=1pt] (char) {\textcolor{white}{#2}};}}
\definecolor{problemblue}{RGB}{100,134,158}
\definecolor{idiomsgreen}{RGB}{0,162,0}
\definecolor{exercisebgblue}{rgb}{0,  .69,  .941}
\definecolor{deepgreen}{rgb}{0.0, 0.5, 0.0}
\definecolor{codegreen}{rgb}{0,0.6,0}
\definecolor{codegray}{rgb}{0.5,0.5,0.5}
\definecolor{codepurple}{rgb}{0.58,0,0.82}
\definecolor{backcolour}{rgb}{0.95,0.95,0.92}
\definecolor{redColor}{RGB}{255,0,0}
\definecolor{Gray}{gray}{0.1}

\lstdefinestyle{mystyle}{
	backgroundcolor=\color{backcolour},   
	commentstyle=\color{codegreen},
	keywordstyle=\color{magenta},
	numberstyle=\tiny\color{codegray},
	stringstyle=\color{codepurple},
	basicstyle=\scriptsize,
	breakatwhitespace=false,         
	breaklines=true,                 
	captionpos=b,                    
	keepspaces=false,                 
	numbers=left,                    
	numbersep=5pt,                  
	showspaces=false,                
	showstringspaces=false,
	showtabs=false,                  
	tabsize=2, numbers=left,
    breaklines=true,
    rulecolor=\color{black}
}
\lstdefinelanguage{test}{%
	language     = python,
	breaklines = true,backgroundcolor=\color{white},escapechar=!,rulecolor=\color{black}, breaklines=true,sensitive=true,  numbersep=5pt, xleftmargin=.015\textwidth, frame=tb,label=test
}
\lstdefinelanguage{source}{%
	language     = python,
	breaklines = true,
firstnumber=0,numberfirstline=false,columns=fullflexible,numbers=left,backgroundcolor=\color{white},
    rulecolor=\color{black}, 
    morecomment=[f][\color{codegreen}]{+\ },
    morecomment=[f][\color{redColor}]{-\ },    
    breaklines=true,sensitive=true, numbersep=5pt, xleftmargin=.015\textwidth, label=test
}
\newcommand*\Suppressnumber{%
  \lst@AddToHook{OnNewLine}{%
    \let\thelstnumber\relax%
     \advance\c@lstnumber-\@ne\relax%
    }%
}

\lstdefinestyle{customc}{
	belowcaptionskip=1\baselineskip,
	breaklines=false,
	frame= single,
	breaklines = true,
	xleftmargin=\parindent,
	language= Python,
	showstringspaces=false,
	basicstyle=\footnotesize\ttfamily,
	keywordstyle=\bfseries\color{green!40!black},
	commentstyle=\itshape\color{purple!40!black},
	identifierstyle=\color{blue},
	stringstyle=\color{codegreen},
	backgroundcolor=\color{gray!4}
}

\DeclareCaptionFormat{listing}{#3}
\newcommand{\killpunct}[1]{}

\begin{document}

%\title{}

\title[A Neuro-Symbolic Compositional Approach for Repository-Level Code Translation and Validation]{\approach: A Neuro-Symbolic Compositional Approach for Repository-Level Code Translation and Validation}

\author{Ali Reza Ibrahimzada}
\orcid{0000-0002-3797-818X}
\affiliation{%
  \institution{University of Illinois Urbana-Champaign}
  \city{Urbana}
  \country{USA}
}
\email{alirezai@illinois.edu}

\author{Kaiyao Ke}
\orcid{0000-0002-0846-1923}
\affiliation{%
  \institution{University of Illinois Urbana-Champaign}
  \city{Urbana}
  \country{USA}
}
\email{kaiyaok2@illinois.edu}

\author{Mrigank Pawagi}
\orcid{0009-0002-6169-4766}
\affiliation{%
  \institution{Indian Institute of Science}
  \city{Bengaluru}
  \country{India}
}
\email{mrigankp@iisc.ac.in}

\author{Muhammad Salman Abid}
\orcid{0000-0002-0135-499X}
\affiliation{%
  \institution{Cornell University}
  \city{Ithaca}
  \country{USA}
}
\email{ma2422@cornell.edu}

\author{Rangeet Pan}
\orcid{0000-0002-8875-1225}
\affiliation{%
  \institution{IBM Research}
  \city{Yorktown Heights}
  \country{USA}
}
\email{Rangeet.Pan@ibm.com}

\author{Saurabh Sinha}
\orcid{0000-0003-4092-2643}
\affiliation{%
  \institution{IBM Research}
  \city{Yorktown Heights}
  \country{USA}
}
\email{sinhas@us.ibm.com}

\author{Reyhaneh Jabbarvand}
\orcid{0000-0002-0668-8526}
\affiliation{%
  \institution{University of Illinois Urbana-Champaign}
  \city{Urbana}
  \country{USA}
}
\email{reyhaneh@illinois.edu}

\renewcommand{\shortauthors}{A. R. Ibrahimzada, K. Ke, M. Pawagi, M. S. Abid, R. Pan, S. Sinha, and R. Jabbarvand}

\begin{abstract}
    Code translation transforms programs from one programming language (PL) to another. One prominent use case is application modernization to enhance maintainability and reliability. Several rule-based transpilers have been designed to automate code translation between different pairs of PLs. However, the rules can become obsolete as the PLs evolve and cannot generalize to other PLs. Recent studies have explored the automation of code translation using Large Language Models (LLMs). One key observation is that such techniques may work well for crafted benchmarks but fail to generalize to the scale and complexity of real-world projects with inter- and intra-class dependencies, custom types, PL-specific features, etc.
We propose \approach, a neuro-symbolic approach to automate \emph{repository-level} code translation. \approach translates both source and test code, and employs multiple levels of validation to ensure the translation \emph{preserves} the functionality of the source program. To break down the problem for LLMs, \approach leverages program analysis to decompose the program into fragments and translates them in the \emph{reverse call order}. 

We leveraged \approach to translate \emph{ten} real-world open-source projects consisting of $\langle$$836$, $8575$, $2719$$\rangle$ \rebuttal{(application and test)} classes, \rebuttal{(application and test)} methods, and \rebuttal{unit} tests. \rebuttal{\approach breaks down these projects into $17874$ fragments} and translates the entire repository. \rebuttal{$96.40\%$} of the translated fragments are syntactically correct, and \approach validates the translations' runtime behavior and functional correctness for \rebuttal{$27.03\%$ and $25.14\%$} of the \rebuttal{application method} fragments. On average, integrated translation and validation takes \rebuttal{$34$} hours (min=\rebuttal{$3$}, max=\rebuttal{$121$}) to translate a project, showing its scalability in practice. For the syntactically or semantically incorrect translations, \approach generates a report including existing translation, stack trace, test errors, or assertion failures. We provided these artifacts to two developers to fix the translation bugs in four projects. They fixed the issues in $20.1$ hours on average ($5.5$ hours for the smallest and $34$ hours for the largest project) and achieved all passing tests. Without \approach, translating and validating such big projects could take weeks, if not months. 
    
\end{abstract}





\maketitle

\section{Introduction}
\label{sec:introduction}

Application modernization offers numerous benefits to developers, including better performance, maintainability, productivity, reliability, and security~\cite{jain2015modernization,khadka2014professionals,khan2022modernization,jamshidi2013cloud}. Manual migration or modernization of real-world projects can be time-consuming and error-prone. Code translation can help automatically convert programs from one programming language (PL) to another. 

Transpilers solely rely on program analysis and perform rule-based translation, failing to translate code between languages that greatly differ in syntax or semantics~\cite{bastidas2023transpilers}. This also makes them very PL-specific; they cannot generalize to newer features of the same PL pairs easily, let alone other PLs.
%\rangeet{One may asks that how our approach support different versions of Java and Python. We can just say that these projects do not support the newer features of PL and cannot be applied...}. 
Finally, the translations lack readability, requiring much effort to understand and validate them, and naturalness, failing to create idiomatic code in the target PL~\cite{pan2024lost}.   
State-of-the-art code translation techniques attempt to harvest the emerging abilities of Large Language Models (LLMs) in code synthesis to overcome the limitations of transpilers~\cite{pan2024lost,yang2024vert,nitin2024spectra}. However, these techniques are still limited to translating simple programs in crafted benchmarks or selected slices of real-world projects due to the following challenges:

\begin{enumerate}[leftmargin=*]
    %\rangeet{Given the latest 128k context window size, would that be a problem. Another solution can be that provide all the dependencies in a single prompt given the large context window. }

    \item \emph{Problem complexity.} The source and target PLs can be fundamentally different in programming paradigms, 
    %(object-oriented vs. procedural), 
    typing,
    %(static vs. dynamic), 
    and memory management. 
    %(manual vs. garbage collection). 
    %The libraries and APIs for one PL may not have an equivalent in another. 
    Some PLs have specific properties that may not exist in others, e.g., constructor overloading in Java. Such complexities are beyond the abilities of existing LLMs to handle, causing them to hallucinate when translating types, code constructs,
    %API usages, 
    or even method names~\cite{pan2024lost}, making translations non-compilable or useless.

    \item \emph{Validation.} The translation should preserve the functionality of the source project. Most existing techniques follow a ``translation first and validation next'' approach, which can postpone the validation and not benefit from the potential use of validation as feedback to correct the translation~\cite{pan2024lost}. A few techniques use formal methods~\cite{yang2024vert,nitin2024spectra} to verify translations on the go. However, these techniques cannot scale to real-world projects. One possible solution for validation is reusing the tests in the source language. However, due to (1) multiple invocations of different methods in unit tests and (2) inherent long call chains in real-world projects, testing a translated method \emph{in isolation} is impossible.  

     \item \emph{Limited context window.} 
     %The prompt and LLM's response are constrained to a specific number of tokens. 
     Concerning repository-level translation, the entire project and, in many cases, even the entire class cannot fit into the context window of current LLMs~\cite{ibrahimzada2024program}. Assuming an unlimited context window, LLMs still suffer from short attention span~\cite{liu2024lost}, preventing them from properly capturing the intra- and inter-procedural dependencies in real-world projects.
\end{enumerate}

This paper presents \approach, a neuro-symbolic\footnote{The keyword symbolic here refers to a general term of symbolic learning in contrast to machine learning and should not be confused with symbolic execution. We refer to combining LLMs and program analysis as a neuro-symbolic approach.} approach for automated repository-level code translation and validation. %\approach decomposes the programs into fragments and translates them by prompting an LLM in the reverse call order. 
%To deal with the complexity of real-world projects, 
\approach leverages static analysis to resolve PL-specific features of the source language (\S\ref{subsec:spt}), decompose the source project into smaller fragments (\S\ref{subsec:decomposition}), and create a compilable project skeleton in the target language (\S\ref{sec:tpsv}). It then starts translating fragments in reverse call order and validates them using existing tests when possible (\S\ref{sec:ictv}). After translating each fragment, \approach updates the project skeleton and ensures the whole project compiles, gradually translating and validating the source project into the target PL. To improve translation quality, static analysis again comes to the rescue: \approach collects relevant context for each fragment, including translated callee methods and surrounding contexts, e.g., class declaration, global variables/fields, etc. It also uses relevant in-context examples based on the specific properties of the fragment to be translated. \approach implements two types of dynamic validation: (1) running the source tests on the translated fragments in isolation using language interoperability (\S\ref{subsubsec:language-interoperability}) and (2) decomposing, translating, and executing the source tests on the translated fragments (\S\ref{subsubsec:test-translation}). Finally, \approach recomposes the translated fragments to create the program in the target PL (\S\ref{subsec:program-recomposition}).

Our approach of compositional translation and validation is PL-agnostic; however, implementing the program transformation component is PL-specific. For the first version of \approach, the implementation supports translating Java code to Python. Our motivations for choosing this PL pair are: (1) Java offers many features that are not supported or common in other PLs by default (e.g., method/constructor overloading, complex types, circular dependencies, local or anonymous inner classes, interfaces, etc.); (2) Python programs are not compiled but interpreted, which makes many translation issues that can be caught at the compile time remain undetected until test execution and challenge the validation; and (3) both PLs are popular (top-5 on the TIOBE index~\cite{tiobe}). 

Using \approach to translate ten real-world Java projects to Python corroborates its \emph{effectiveness}: \rebuttal{It can translate $17874$ field/method/test fragments, with $96.40\%$ syntactic correctness. For the $4654$ application method fragments that can be further evaluated through test execution, \approach achieves $27.03\%$ runtime validation and $25.14\%$ functional equivalence using the source tests}. \approach is \emph{scalable}, completing translations in \rebuttal{$34$} hours, on average. Human subjects improved partial translation of \approach and achieved passing test suites within $20.1$ hours, on average, showing \emph{practicality} of \approach. These results were achieved using a moderate-size open-access LLM (\llm~\cite{guo2024deepseek}). \rebuttal{A stronger model, i.e., GPT-4o, improves the performance of \approach to $99.2\%$ syntactic correctness for all fragments and $27.95\%$ functional equivalence for method fragments, with an overhead of $\$14.39$ per project, on average. The affordable cost is due to the novel features of the pipeline, namely, decomposition into fragments, prompt crafting, in-isolation validation of translations, and efficient feedback loop.} 
% \reyhan{for the gpt, please report the results for that model not the combined one}

%, and using bigger/stronger models such as GPT-4~\cite{openai2023gpt4} or Claude-3~\cite{claude} will improve the results. \reyhan{the numbers in this paragraph should be updated in the final pass}

To the best of our knowledge, \approach is the \emph{first technique to translate an entire repository}, including tests, and generates validated translations (considering existing tests). The only prior repository-level translation attempt using GPT-4~\cite{pan2024lost} (translating Apache Commons CLI from Java to Python) resulted in non-compilable code, let alone the translation being validated. \approach is also \emph{the first technique leveraging language interoperability for in-isolation validation of translated fragments}. 
%Compared to the only repository-level translation attempt using GPT-4~\cite{pan2024lost} (translating Apache Commons CLI from Java to Python) that resulted in non-compilable code, let alone the translation being validated, \approach translated the entire repository, producing \rebuttal{$98.9\%$} syntactically correct and \rebuttal{$70.7\%$} functionally validated code. 
The effort of human subjects to fix translation bugs by \approach and achieve passing tests creates pragmatic bug data sets for testing, bug localization, and program repair research. Our code and artifacts are publicly available for reproducing the results or translating new projects~\cite{website}. 

\vspace{-10pt}

\section{Challenges in Repository-Level Code Translation}
\label{sec:illustrative-example}

\begin{figure*}
    \centering
    \includegraphics[width=0.97\textwidth]{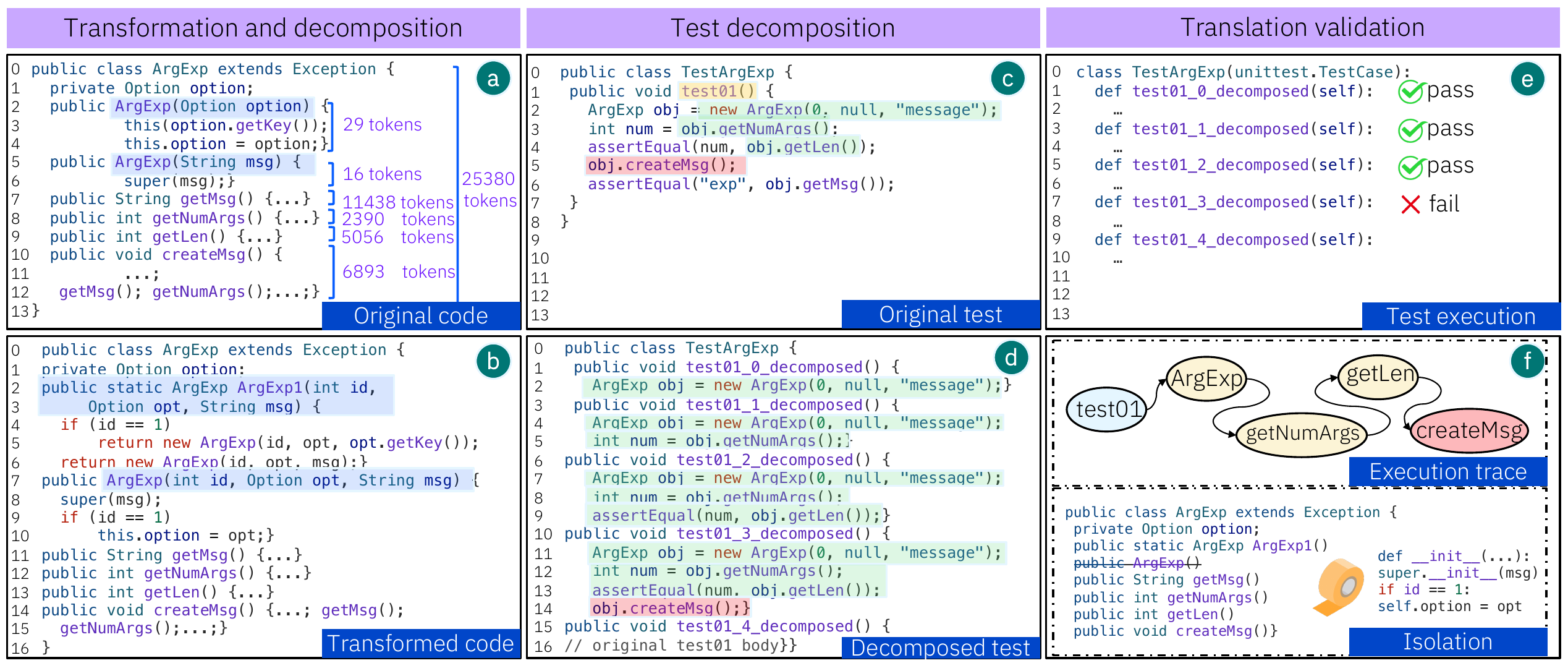}
    \vspace{-10pt}
    \caption{Illustration of key challenges in repository-level code translation and \approach addressing them.}
    % \ali{reyhan wants test02 to be removed.}
    \vspace{-15pt}
    \label{fig:illustrative-example}
\end{figure*}

To illustrate the most notable challenges in repository-level code translation and validation, we use the hypothetical example in Figure~\ref{fig:illustrative-example}, inspired by the complexities in real-world Java projects. 

\textbf{Challenge 1: Class Size.} 
The class consists of $25,380$ tokens~(\circled[fill=teal]{a}). 
%\saurabh{We should say ``for illustrative purposes'' as the actual MissingArgumentException class is very small (10 lines)}. 
Instructions for translating the code, in-context examples, and the model's response can also significantly increase the number of input tokens. While some commercial LLMs support tens of thousands of tokens, many open-access LLMs do not. For example, \llm~\cite{guo2024deepseek} used in this paper has a context window of $16,384$ tokens, of which only $4,096$ tokens can be used for generation. To address this challenge, \approach decomposes Java application classes into smaller \emph{field} and \emph{method fragments} and translates each separately in reverse call order (\S\ref{subsec:spd}, \S\ref{subsec:td}).
% thus, breaking down a complex translation task into simpler, more-manageable tasks.

\textbf{Challenge 2: PL-specific Properties.} 
Java programs frequently use method/constructor overloading, which is not supported directly in Python~(\circled[fill=teal]{a}). This example shows instances of constructor overloading (lines 2 and 5). In Python, declaring two constructors is allowed; however, at runtime, the last declaration overrides all previous constructors, resulting in unexpected behavior.
% \reyhan{@Ali, explain what happens in python if you don't address this}.
To address this issue, \approach employs program analysis to refactor the original code while preserving the functionality (through test execution). The transformation includes changing the constructor's name, updating the references, and changing the constructor's implementation. The transformed code~(\circled[fill=teal]{b}) makes the source program amenable to translation to Python. 

\textbf{Challenge 3: Validation.}
To illustrate the challenges with validation, consider \texttt{\small test01}~(\circled[fill=teal]{c}) that invokes \emph{four} methods in its body (\texttt{\small{ArgExp}}, \texttt{\small{getNumArgs}}, \texttt{\small{getLen}}, and \texttt{\small{createMsg}}) to test the functionality of method \texttt{\small{getMsg}} in the assert statement. Suppose we can successfully translate all methods except \texttt{\small createMsg}. If we choose test translation (a natural way of validating code translation), the execution of the translated test results in a runtime error when invoking \texttt{\small createMsg}. As a result, a translation issue in one method casts a shadow in validating the translation of the other methods. We refer to this issue as the \emph{test translation coupling effect}. To overcome this challenge, \approach executes source language tests as-is (i.e., without translation) by leveraging a language-interoperability framework called GraalVM~\cite{graalvm} (\circled[fill=teal]{f}). In this setting, a test in the source language is executed each time one of its invoked application methods (method fragments) is translated. This approach validates \emph{functional equivalence} of each method \emph{in isolation} as the other invoked application methods during test execution remain in the source language. %This approach provides test execution \emph{in isolation} for each translated application method to validate it while being insulated from buggy translations of its dependencies.

\textbf{Challenge 4: Test Translation.}
GraalVM has certain limitations (\S\ref{subsubsec:language-interoperability}), which prevents \approach from validating all the code fragments in isolation. Furthermore, we need to translate tests regardless of whether they are used for validation to maintain the translated projects in the target language. Test errors due to test translation coupling effect under-approximate the quality of translation: failing to validate the translation of four methods because of one incorrect translation.
% Root causing the translation bugs also requires additional efforts from developers, i.e., looking at the stack trace and coverage.
To overcome this challenge, \approach decomposes the original test suite into \emph{test fragments} (\circled[fill=teal]{d}). Executing the translated decomposed test suite results in three test passes (\circled[fill=teal]{e}), validating the \emph{runtime behavior} of three methods that the original test suite could not promptly provide. 

An alternative approach is parsing the stack trace and code coverage results for each runtime error during translation. However, test decomposition is a cleaner way to see the results per test execution promptly. It is also done once before translation. In translation to interpreted languages such as Python, specifically, the execution of test fragments can validate the runtime behavior of methods before waiting for functional validation. For fragments that GraalVM cannot validate, if \approach can successfully translate all the methods invoked during test execution and the test passes, such a test will also be used for validating \emph{functional correctness}. 

\section{Overview of Approach}
\label{sec:approach-overview}

\begin{figure*}[t]
    \includegraphics[width=0.85\textwidth]{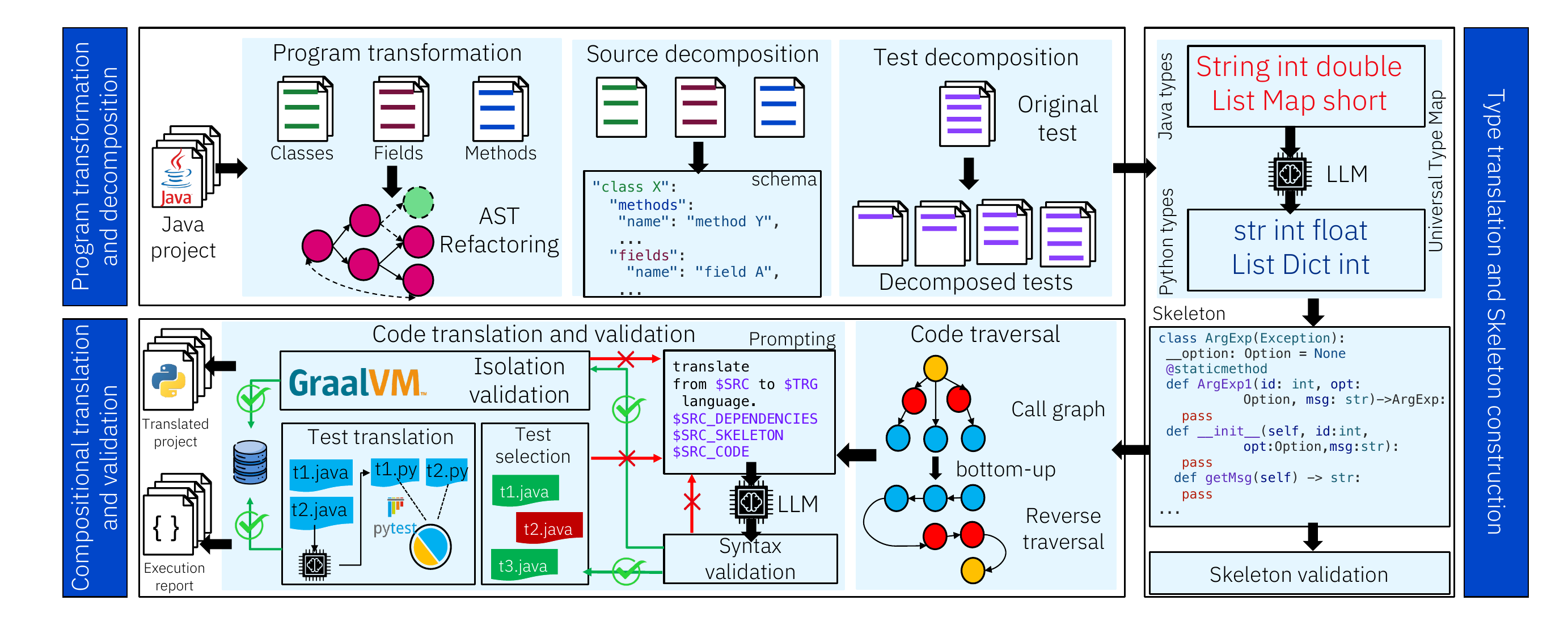}
    \vspace{-15pt}
    \caption{Overview of \approach.}
    \label{fig:framework}
    \vspace{-25pt}
\end{figure*}

\approach consists of three main phases, shown in Figure~\ref{fig:framework}: program transformation and decomposition (\S\ref{sec:sptd}), type translation and skeleton construction (\S\ref{sec:tpsv}), and compositional translation and validation (\S\ref{sec:ictv}). The first two phases aim to decompose and simplify the repository-level code translation problem for LLMs, helping the third phase yield high-quality validated translations.  

The program transformation and decomposition phase first refactors the PL-specific properties of the source program into programming paradigms common among many PLs (\S\ref{subsec:spt}). Next, it decomposes the source project into smaller units, i.e., \textit{fragments}, and stores fragment dependencies in a data structure called \textit{schema} (\S\ref{subsec:decomposition}).

The type translation and skeleton construction phase takes the schema as input and produces \emph{target project skeleton}, i.e., a compilable project in the target language with method signatures but no method implementation (\S\ref{subsec:scv}). The first translation step happens here, where the source PL types are translated to the target PL to ensure that class skeletons are compilable (\S\ref{subsec:typer}). The outcome of type translation is a type mapping from the source to the target PL, which \approach can reuse in translating other projects.

The compositional translation and validation phase takes the schema and project skeleton as inputs and translates fragments, in reverse call order, 
%from the source to the target PL 
by prompting an LLM. After translating a fragment, it updates the class skeleton with a new translation and checks whether the skeleton compiles. For a method fragment, \approach looks for corresponding tests and, if any exist, uses them to validate the fragment. The first level of validation is performed through GraalVM's language interoperability to isolate the validation of the method using tests in the source language. Next, \approach translates and executes the tests. In case of compilation errors or test failures, \approach reprompts the LLM with feedback (from the compilation/runtime errors) to improve the translation. If no improvement is achieved within a certain budget, \approach continues to the next fragment until all are translated. For methods whose translations are not compilable or result in test errors/failures, \approach generates reports consisting of existing translations and relevant artifacts, such as stack traces, test errors/failures, and test coverage information.

\vspace{-5pt}
\section{Program Transformation and Decomposition}
\label{sec:sptd}

\subsection{Program Transformation}
\label{subsec:spt}

This component performs semantics-preserving refactoring of method and constructor overloading in Java code to make it amenable to translation to Python.  
%languages, such as Python, that do not support method overloading. 
Other Java-specific features, namely, circular dependencies, inner classes, interfaces, and abstract classes, are handled while constructing the project skeleton in Python (\S\ref{subsec:scv}). The reason for resolving method and constructor overloading in the source language is that we have to change the implementation, i.e., call sites to methods and constructors. Therefore, such changes should be validated using source tests before translation. 

\begin{figure*}[t]
    \includegraphics[width=1\textwidth]{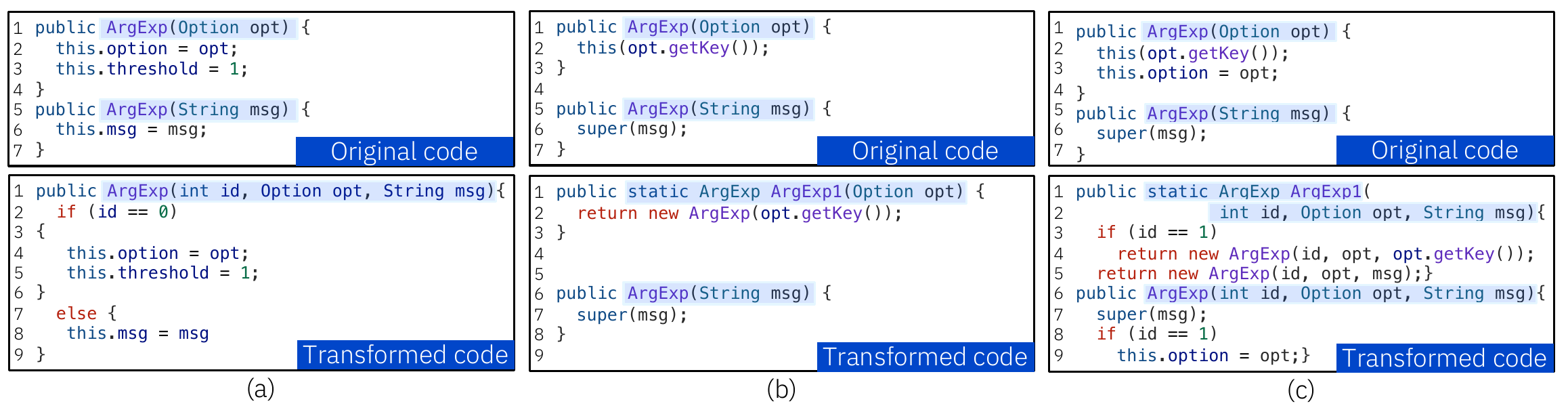}
    \vspace{-25pt}
    \caption{Constructor overloading patterns and their corresponding transformations.}
    % \reyhan{change the order to be (c), (b), and (a)}
    \vspace{-13pt}
    \label{fig:transformation}
\end{figure*}

For overloaded methods, \approach makes each method name unique by adding an integer suffix (starting at $0$) to the name, and updates all call sites based on the new method names. Resolving overloaded constructors is not as straightforward, as they should have the same name as the enclosed declaring class. Furthermore, the invocation of constructors inside each other and the Java inheritance mechanism makes constructor overloading complex. Our algorithm (Algorithm~\ref{alg:transformation}) for resolving the constructor overloading handles three prominent patterns shown in Figure~\ref{fig:transformation}.\footnote{Following the best practices for constructor overloading from \textit{Stack Overflow} and analyzing the use of constructor overloading in open-source projects, we categorized the use cases into the three patterns.}

\begin{wrapfigure}{R}{0.5\linewidth}
  \centering
  \scriptsize
  \begin{minipage}{\linewidth}
  \vspace{-0.4cm} % to remove the whitespace above algorithm
\begin{algorithm}[H]
    \input{Resources/Algorithms/transformation}
\end{algorithm}
  \vspace{-0.35cm} % to remove the whitespace below algorithm
\end{minipage}
\end{wrapfigure}

The first pattern (Figure~\ref{fig:transformation}-a) shows multiple independent constructors. \approach merges these constructors into one and uses an \texttt{\small id} parameter to differentiate between them. All call sites of the constructors are updated accordingly to use the appropriate \texttt{\small id} value. 
The second pattern (Figure~\ref{fig:transformation}-b) involves a constructor call chain using \texttt{\small this()}. \approach transforms the first constructor into a factory method and invokes the second constructor inside it. Factory methods are static, and \approach updates the call sites to invoke them directly on the class, e.g., \texttt{\small ArgExp.ArgExp1(id,opt,msg)}. 
The last pattern (Figure~\ref{fig:transformation}-c) is similar to the second one, except that both constructors implement some code. \approach refactors the first constructor into a factory method and adds an extra \texttt{\small id} parameter to differentiate between behaviors implemented by different constructors. The constructor call sites are updated accordingly, as in the previous cases. Real-world projects often combine these patterns, which \approach handles using Algorithm~\ref{alg:transformation}. 

\subsection{Program Decomposition}
\label{subsec:decomposition}
Translating the entire repository of real-world projects is a very complex problem. As a result, \approach breaks down projects into fragments, performs the translation and validation at the fragment level, and re-composes the translation as a repository in the target language.  

\subsubsection{Source Decomposition}
\label{subsec:spd}
Real-world projects can include hundreds of files with thousands of lines of code, which exceed the context window of state-of-the-art LLMs. \approach employs static analysis to decompose code into smaller fragments, i.e., \emph{field fragments} and \emph{method fragments}. A field fragment includes modifiers, type, name, and field value. A field fragment can belong to an application or test class. A method fragment includes the method signature and can be an application or test method (e.g., helper methods or unit tests). During decomposition, \approach extracts meta-information related to the fragments, such as their location (e.g., start and end line), code (e.g., implementation between start and end line), dependencies (e.g., callers and callees), types (of inputs and output), and other necessary information such as file paths, class inheritance, imports, and method annotations. \approach stores fragments and their corresponding collected meta-data in a data structure called \emph{schema}, which is used in the other phases. \approach also extracts the call graph to guide the translation, i.e., to translate fragments in reverse call order.

\vspace{-4pt}
\subsubsection{Test Decomposition}
\label{subsec:td}

The burden of checking \emph{functional equivalence} in \approach is on GraalVM. Yet, we still need to translate and execute tests to validate the fragments that cannot be validated by GraalVM (\S\ref{subsubsec:language-interoperability}). Unit tests in real-world projects can invoke multiple methods and include multiple assert statements. Furthermore, long call chains are inevitable in real-world projects due to the high degree of intra- and iter-procedural dependencies. As we show later (\S\ref{sec:rq-test-decomposition}), the average number of direct method invocations and method executions in tests for our studies subjects are $3$ and $27$, respectively. This can result in \emph{test translation coupling effect}, discussed in \S\ref{sec:illustrative-example}. 

To enable test translation for runtime validation or checking functional equivalence, \approach decomposes each unit test into a series of \emph{test fragments}, as shown in Figure~\ref{fig:illustrative-example}-d. It uses each statement with a call to an application method as a cutting point. For statements enclosed by branches, loops, of exception-handling blocks, \approach includes the entire block. This process generates an ordering of executable test fragments for each unit test. Each test fragment includes all the statements of the lower-order fragments, along with additional statements that invoke one additional method not invoked by previous fragments. \approach executes test fragments in increasing order until a test fails and skips running following fragments, as they will also fail.

\section{Type Translation and Skeleton Construction}
\label{sec:tpsv}

\subsection{Type Translation}
\label{subsec:typer}

\begin{wrapfigure}{R}{0.38\linewidth}
  \centering
  \scriptsize
  \vspace{-15pt}
  \begin{minipage}{\linewidth}
  \vspace{-0.7cm} % to remove the whitespace above
    \includegraphics[width=\linewidth]{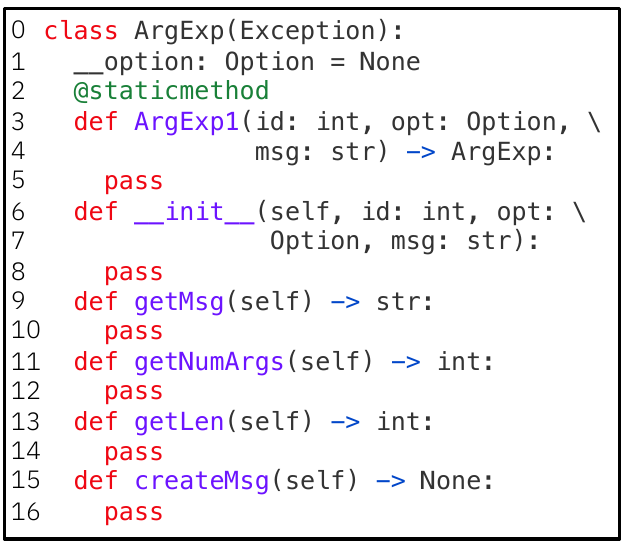}
    \vspace{-18pt}
    \caption{Target PL skeleton for the example in Figure~\ref{fig:illustrative-example}-b.}
    \label{fig:skeleton}
  \vspace{-0.8cm} % to remove the whitespace below
\end{minipage}
\end{wrapfigure}
Automatically resolving types is a challenging problem~\cite{terekhov2000realities,guizzo2023MBTA}, and a large body of work has attempted to address this, mostly using symbolic rule-based approaches~\cite{qiu1999pltranslation,dony2010interconversion,broggi2023interoperability,roziere2020unsupervised,coco2018JPT,lano2024modeldriven}. \approach employs a Retrieval-Augment Generation (RAG)~\cite{lewis2020retrieval} technique for finding equivalent types in the target language. To that end, it first extracts all the types in the source language of a given project. Custom application types are resolved during the translation as \approach translates fragments and classes in the target language. For the remaining types, it crawls the API documentation of the source language and retrieves the relevant description of each type. 

To form the prompt,\footnote{Due to space limit, we omit the prompt; please refer to our artifacts~\cite{website} to see the prompts used for type resolution.} \approach uses the retrieved description and instructs the model with an in-context example to return the most relevant type in the target language, given the use of types in the source language and the retrieved description. To account for potential hallucination in LLM's response, i.e., returning a type that does not exist in Python, \approach employs a simple Python script, uses the translated type as an annotation, executes the script, and keeps the ones that have no syntactic or runtime issue. The types in the source language and their corresponding in the target language form a data structure called \emph{universal type mapping}. In practice, \approach reuses or augments the mapping when translating new projects. 

\vspace{-5pt}
\subsection{Skeleton Construction}
\label{subsec:scv}

\approach builds the project's structure in the target language before translation. This step is necessary for compositional translation and validation, as \approach can insert the translated fragments into the project, compile it, or even execute the existing translated test suites, gradually completing the translation. At this step, \approach also resolves Java-specific features in Python before starting the translation. Specifically, it resolves circular imports and dependencies, inner classes, interfaces, and abstract classes. Figure~\ref{fig:skeleton} shows the class skeleton corresponding to the illustrative example of Figure~\ref{fig:illustrative-example}-b. 

At the first step, \approach creates Python classes corresponding to each application class in Java. The fields in the classes are set to \texttt{\small None}, and \approach uses the information in schema (\S\ref{subsec:spd}) to ensure the naming corresponds to the type of their access modifier in the source language. In the example of Figure~\ref{fig:skeleton}, the translation of field \texttt{\small private Option option;} in Java is \texttt{\small \_\_option: Option = None}. The classes also include method signatures, with their body set to \texttt{\small pass}. \approach uses the universal type mapping (\S\ref{subsec:typer}) to create relevant types in the method signature. Once the initial skeleton is created, \approach leverages the extracted call graph during program decomposition to detect circular dependencies. If such dependencies exist, \approach resolves them with local imports. For inner classes, \approach unfolds them in Python and uses \emph{dot notation} to access specific methods and fields (e.g., \texttt{\small Class.methodName}). Finally, \approach implements best practices in Python and subclasses all abstract classes and interfaces from \texttt{\small abc.ABC} class. Here, \texttt{ABC} is a class from the \texttt{\small abc} module in the Python standard library, which is used for defining abstract base classes.
% \vspace{-20pt}
\section{Compositional Translation and Validation}
\label{sec:ictv}

\begin{wrapfigure}[14]{r}{0.45\textwidth}
    \centering
    \vspace{-10pt}
    \includegraphics[width=\linewidth]{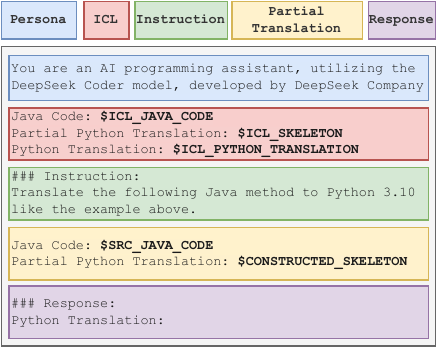}
    \vspace{-18pt}
    \caption{Prompt structure in \approach.}
    \label{fig:prompt}
    %\vspace{1cm} % to add the whitespace below
\end{wrapfigure}
\approach translates method fragments in reverse call order. \rebuttal{It takes the project's  call graph, removes the back edges to make it acyclic, computes topological order (i.e., linear ordering of its vertices), and translates method fragments (corresponding to vertices) in reverse topological order. Field fragments are independent; therefore, AlphaTrans translates them before method fragments.} The algorithm takes a fragment $F$, LLM $M$, Skeleton $S,$ and a series of parameters as inputs, translates the fragment, recomposes the skeleton with the successfully translated fragment, and returns translation outcomes: (1) syntax check (denoted by the ``non-parseable'' and ``parseable'' labels), (2) functional equivalence check (denoted by the ``graal-fail'', ``graal-success'', and ``graal-error" labels), and (3) translated test execution check (denoted by ``not-exercised'', ``test-fail'', and ``test-success" labels).

\approach employs iterative and feedback-based prompting. That is, if one of the mentioned checks fails, e.g., the translated fragment is not syntactically correct, it prompts the model for another translation. To control the number of iterations, \approach uses a reprompting budget (\textit{repromptBudget}). The algorithm takes the minimum ($min_{budget}$) and maximum ($max_{budget}$) values for the budget and dynamically sets reprompting budget to a number between the lower and upper bounds based on coverage information, e.g., the budget is close to $max_{budget}$ for a fragment if it is exercised multiple times (high hit rate based on coverage information) by different unit tests. The rationale is to give more importance to fragments covered by more tests to eventually increase translation validation success.

\begin{wrapfigure}{R}{0.5\linewidth}
  \centering
  \scriptsize
  \begin{minipage}{\linewidth}
%\vspace{-0.4cm} % to remove the whitespace above algorithm
\begin{algorithm}[H]
    \input{Resources/Algorithms/comp-translation-validation}
\end{algorithm}
\vspace{-0.4cm} % to remove the whitespace below algorithm
\end{minipage}
\end{wrapfigure}

Algorithm~\ref{alg:translation-validation} shows the main translation and validation loop (lines~3--31), which runs until the reprompting budget is exhausted. Inside the loop, \approach first crafts a unique prompt based on the template shown in Figure~\ref{fig:prompt} and then instructs the LLM to translate the fragment (lines~4--5). It then validates the generated translation in multiple steps. The first step checks for syntactic correctness and assigns proper labels to $\mathit{TVO}$ (lines~6--10). Then, \approach leverages GraalVM for isolation-based validation of fragment $F$ (lines~12--20), if there exists a test in the source language covering the fragment during its execution. 

Finally, it translates and executes decomposed fragment tests: if there are no eligible tests (a test becomes eligible if all its dependencies are translated) for the fragment, \approach simply assigns the ``\emph{not-exercised}'' label to the fragment and moves on to the next one (lines~21--23). Otherwise, it translates the tests, executes them to validate the fragment, and assigns test outcome labels in \textit{TVO} (lines~24--31). In case of a test failure, \approach extracts all involved fragments and reprompts them with feedback extracted from test execution. 

Due to inherent intra- and inter-procedural dependencies in real-world projects, the number of fragments involved in reprompting could be high, logarithmically increasing the translation time. \approach filters out those with GraalVM label ``\emph{graal-success}'', ranks the remaining based on suspiciousness score, and reprompts $topK$ suspicious fragments. The suspiciousness score for fragments is calculated such that a fragment with more failing tests will get a higher score and, therefore, ranked higher among other fragments.
% \reyhan{add explanation about how you compute the susp score}

Our prompt template consists of \emph{five} distinct parts (Figure~\ref{fig:prompt}). The first part is the \emph{persona} message used by \llm during instruction fine-tuning and is required for producing the best outputs. The next part introduces the \emph{In-Context Learning (ICL)} example, which reflects the complexities of code translation and instructs LLM on how to deal with them. %The ICL example also forces the model not to generate junk in its response. 
The green part shows the natural language instruction given to the model. After describing the objective, the prompt embeds the source Java code along with \emph{partial translation} as a skeleton, which includes all dependencies and translations of the fragments invoked by the current one. The prompt concludes with \texttt{\small \#\#\# Response:} keyword to guide the model for code generation.

%\vspace{-5pt}
\subsection{Language Interoperability}
\label{subsubsec:language-interoperability}

GraalVM~\cite{wurthinger2013one,graalvm} is a Java Development Kit by Oracle. It offers the \emph{Polyglot} API~\cite{polyglotapi}, which allows the integration of programs written in different guest languages within a Java-based host application. In the context of this work, GraalVM allows execution of Python code from Java and vice versa~\cite{abid2024gluetest}. \approach leverages the Polyglot API to perform \textit{in-isolation} validation of the fragments by replacing the Java implementation of a method with its translated Python version while keeping the rest of the project in Java. It then executes the Java tests covering the fragment to validate the functional equivalence of the translation. 

The Polyglot API allows access to Python data objects from Java and vice-versa, as these objects reside in a shared memory space. However, objects must be cast to appropriate types for passing parameters to, and processing returned values from, polyglot calls. 
The Polyglot API can perform this casting implicitly for only a few simple data types. \approach builds on top of the Polyglot API to provide a framework to create a Python program state that is isomorphic to the Java program state. The Python translation is restricted to this isomorphic state, and the states are synchronized after method calls to preserve the isomorphism. \approach allows for the casting of user-defined types as well as several built-in and library types. Using both static and dynamic type information of Java objects, \approach can disambiguate target types when casting Python objects to Java types. It further preserves object identities and aliasing during such casting and propagates exceptions across language boundaries.

To validate the translation of a method $m$ in isolation, \approach creates an instrumented version of the Java source code. We refer to this instrumented Java project as the \textit{primal project}, $P_m$. During instrumentation, \approach replaces the original Java implementation, $m_J$ of $m$ with a polyglot call to its Python implementation, $m_P$. $m_P$ resides inside a Python project, which we refer to as the \textit{dual project}, $D_m$. The structure of $D_m$ is similar to that of the original Java project. All other methods in $D_m$ wrap a call to the corresponding methods in $P_m$. Doing so provides an interface for $m_P$ to execute with access to all other methods and fields, although these are defined only in $P_m$. Using the call graph for the Java project, \approach determines all test methods that invoke $m$ and executes them in $P_m$ to validate the translation $m_P$.
This \textit{in-isolation} validation approach is limited in the sense that it can handle only a limited number of built-in and library types. In certain cases, such as reference cycles involving maps or objects with impure methods for hashing, the isomorphism between Java and Python states may not be maintained. Furthermore, it may sometimes not be possible to disambiguate target types when casting Python objects to Java types, for example, if the target object has type \java{List<Object>}.

\vspace{-5pt}
\subsection{Target Program Recomposition}
\label{subsec:program-recomposition}

%When validating application method fragments with test translation and execution, 
\rebuttal{\approach recomposes the project skeleton with syntactically correct translated fragments at each iteration of Algorithm 1, gradually constructing the project in the target PL. The recomposed target program in Python is executed against eligible translated tests, i.e., those that cover only the translated fragments.}
%Depending on test execution results, AlphaTrans either keeps the translation or replaces the fragment body with \texttt{\small{pass}} to move on to the next fragment. }

%\rebuttal{Specifically, project skeletons are stored in structured JSON files where each fragment is identified by its unique ID. We use these IDs to locate the translated fragments and update skeletons in JSON files by read/write operations. 
%The pipeline then combines class skeletons with the body of the translated fragment.
%and creates stand-alone Python files. 
%Construction of the project skeleton in Python and updating it upon each fragment translation enables AlphaTrans to migrate the code step by step (per each fragment translation) and validate incremental project translation through test execution.

\vspace{-5pt}
\subsection{Test Translation}
\label{subsubsec:test-translation}

Similar to translating application fragments, \approach also translates test fragments. Using the dependency information captured during static analysis, it crafts prompts for unit tests along with their dependencies for the model to translate. The ICL examples used for prompting test fragments differ from prompts used for translating application method fragments. The focus of ICL examples here is to prevent the LLM from hallucinating the used assert statements in the source PL to the target PL. To construct ICL examples for test fragment translation, we created a pool of in-context examples, where each example shows the Python assert statements equivalent to Java assert statements in the context of a test. When prompting a test fragment, \approach detects the assert statement in the fragment and retrieves the corresponding examples from the pool. 
%\approach crawls the API documentation of JUnit~\cite{junit} assert calls, and dynamically create in-context examples, showing the assert call counter-parts in Python. 
For translated tests, only syntactic validation is performed as there is no other means of validating their translations. 
%The translated tests in Python are then recomposed into a Pytest~\cite{pytest} test suite, which \approach uses to execute for validating runtime behavior

\section{Empirical Evaluation}
\label{sec:evaluation}

To evaluate different aspects of \approach, we investigate the following research questions:

\begin{enumerate}[label=\bfseries RQ\arabic*:]
  \item \textit{Effectiveness of \approach}. To what extent \approach can automatically resolve types from source to target PL? Can \approach effectively translate real-world projects? 

  \item \textit{Translation Bugs and Fixes}. How much effort do developers spend completing the partial translations by \approach? 
  %and achieving all passing tests? 
  What is the nature of translation bugs? 

   \item \textit{Impact of Test Decomposition}. What is the impact of test decomposition on validation results? 
  
  \item \textit{Impact of Test Coverage}. To what extent does a test suite with higher coverage impact the validation results?

  \item \rebuttal{\textit{Ablation Study}. To what extent do program transformation, choice of LLM, and program decomposition impact the performance of \approach?}
\end{enumerate}

\vspace{-10pt}
\subsection{Experiment Setup}
\label{sec:experimental-setup}

%\subsubsection{Subjects:} 
We followed three steps for selecting subjects: \textbf{1- Mining:} We mined GitHub and retrieved a list of repositories that use Java as the primary language, are self-contained (include build files, etc.), and have more than $30$ stars with at least one commit pushed within the last $12$ months.
\textbf{2- Filtering:} We filtered out projects based on the number of call edges in their call graphs: removed those with less than $2,000$ call edges to ensure the subject projects are big enough to challenge \approach. We also removed those with more than $ 30,000$ call edges to reduce the computation and manual effort for further steps. Per GraalVM requirements, we only selected projects we could successfully build (compile and achieve green tests) using Java at $21$. 
\textbf{3- Reduction:} \sloppy \approach currently supports the following Java APIs: core Java API (\texttt{\small java.util}, \texttt{\small java.text}, \texttt{\small java.lang}, \texttt{\small java.io}, \texttt{\small java.nio}, \texttt{\small java.net}, \texttt{\small java.time}, and \texttt{\small java.math}) and third-party libraries (\texttt{\small org.opentest4j}, \texttt{\small org.slf4j.Logger}, and \texttt{\small org.junit}). We automatically removed all other third-party library dependencies and their usage in the source code in the selected projects. We chose a project if at least $50\%$ of its total methods were preserved after such process. Table~\ref{table:stats} shows the list of ten projects used in the evaluation of \approach and details about their size (classes, methods, tests, and fragments). 

%\subsubsection{Program analysis:} 
\approach uses CodeQL~\cite{codeql} and tree-sitter~\cite{tree-sitter} for static analysis. For running tests, validating translation, and computing coverage, \approach uses GraalVM $21.0.3+7.1$~\cite{graalvm}, JUnit 4 and 5~\cite{junit}, Pytest 8.2.1~\cite{pytest}, JaCoCo~\cite{jacoco}, and Python's \texttt{\small coverage}~\cite{coverage-lib}. 
%We modified Python code coverage implementation for statement coverage to compute method coverage. 
%\approach uses GraalVM $21.0.3+7.1$~\cite{graalvm} with Java $21$.
%for isolation-based translation validation.
%\subsubsection{LLM:} 
\approach works with API- and open-access LLMs. 
%We considered \gptfo (one of the best-performing commercial LLMs) and \llm (a moderate-size LLM outperforming many other open-source models) as LLM components of \approach for the evaluation.}
We considered the following criteria for selecting the LLM: (1) for better reproducibility, we prioritized open-access models; (2) due to computing constraints, we wanted an LLM with moderate size ($>10B$ but $<70B$ parameters); (3) the model should perform reasonably well in code-related tasks; and (4) the model should have fast inference time due to the huge number of prompts. Per the mentioned criteria, we selected \llm~\cite{guo2024deepseek} for the \rebuttal{main} experiments \rebuttal{(RQ1--RQ4)}. \rebuttal{We also used \gptfo, one of the best-performing commercial models, for RQ5 (\S\ref{sec:rq-ablation}) to demonstrate the impact of stronger models on improving the performance of \approach.} 
We prompted \rebuttal{models} with the temperature set to~$0$ for reproducibility and used their default settings for other parameters. For the base prompting (Algorithm~\ref{alg:translation-validation}), we set the minimum and maximum values of the reprompting budget to $3$ and $5$. For the feedback prompting, we set the reprompting budget to $1$, i.e., \approach attempts to fix issues with feedback only once. 
\subsection{RQ1: Effectiveness of \approach}
\label{sec:rq-effectiveness}

In this RQ, we evaluate \approach in (1) type translation and skeleton construction (\S\ref{subsubsec:rq-effectiveness-preprocessing}) and (2) compositional translation and validation (\S\ref{subsubsec:rq-effectiveness-translation}).

\subsubsection{Effectiveness in Type Resolution and Skeleton Validation:}
\label{subsubsec:rq-effectiveness-preprocessing}

\approach extracted $1,797$ distinct types from the source projects and attempted to translate them to equivalent Python types. Of these, $915$ are application types (i.e., classes defined within the Java projects) and were directly resolved during skeleton construction. \approach prompts \llm to resolve the remaining $882$ and successfully translated $738~((915+738)/1,797=91.99\%)$ of them: the generated types passed the syntactic and runtime check. The column \emph{ATR} in Table~\ref{table:stats} shows the results of automated type resolution. Because type resolution is essential to project skeleton construction, we manually checked the type mappings generated by \llm and also translated the $144$ unresolved types.

Through manual investigation of the automatically resolved types, we observed that \llm's type resolution for $182$ cases, while \emph{correct}, can be \emph{improved}. For example, \approach translated  \texttt{\small java.io.File}, a class concerning file manipulation functionality to \texttt{\small str}. The resolved type can represent file paths in Python but lacks features for file manipulation. We suspect this translation is impacted by the Java use case provided in the prompt. While this translation is correct for the use case, we replaced it with \texttt{\small pathlib.Path} to have a more generic type mapping.
We also augmented the type mapping with additional types in the target language for $38$ types. For example, \approach translated \texttt{\small java.nio.Buffer} to \texttt{\small bytearray}, which is correct as they both provide a mutable sequence of bytes with efficient in-place modifications. However, \texttt{\small array.array} and \texttt{\small memoryview} also provide similar functionality with efficient and low-level data manipulation capabilities. Consequently, we augmented type mapping to \texttt{\small typing.Union[bytearray, array.array, memoryview]}. Given that type mapping can be reused, this one-time manual effort increases the chance of \approach's success on unseen projects.  
\begin{table}[t]
    \setlength{\tabcolsep}{1pt}
    \scriptsize
    \centering
    %\vspace{-10pt}
    \caption{Effectiveness of \approach in program transformation, automated type translation, and skeleton validation. \textbf{ATR:} Automated Types Resolution, \textbf{SV:} Skeleton Validation.
    \rebuttal{The number of classes and methods include both application and test classes/methods.}}
    \vspace{-5pt}
    \begin{tabular}{l|cccc|c|c|cccc}
\hline
                                    &                                       &                                       &                                                                                        &                                                                                           &                                     &                                    & \multicolumn{4}{c}{\textbf{\# Fragments}}                                                                                                                                                                                                                                          \\ \cline{8-11} 
                                    &                                       &                                       &                                                                                        &                                                                                           &                                     &                                    & \multicolumn{2}{c|}{\textbf{Fields}}                                           & \multicolumn{1}{c|}{}                                                                                         &                                                                                   \\ \cline{8-9}
\multirow{-3}{*}{\textbf{Subjects}} & \multirow{-3}{*}{\textbf{\# Classes}} & \multirow{-3}{*}{\textbf{\# Methods}} & \multirow{-3}{*}{\textbf{\begin{tabular}[c]{@{}c@{}}\# JUnit\\ Tests\end{tabular}}} & \multirow{-3}{*}{\textbf{\begin{tabular}[c]{@{}c@{}}Method\\ Coverage (\%)\end{tabular}}} & \multirow{-3}{*}{\textbf{ATR (\%)}} & \multirow{-3}{*}{\textbf{SV (\%)}} & \multicolumn{1}{l|}{\textbf{Application}} & \multicolumn{1}{l|}{\textbf{Test}} & \multicolumn{1}{c|}{\multirow{-2}{*}{\textbf{\begin{tabular}[c]{@{}c@{}}Application\\ Methods\end{tabular}}}} & \multirow{-2}{*}{\textbf{\begin{tabular}[c]{@{}c@{}}Test\\ Methods\end{tabular}}} \\ \hline
cli~\cite{commons-cli}                                 & 58                                    & 664                                   & 437                                                                                    & {\color[HTML]{000000} 94.14}                                                              & 96.60                               & 100                                & 104                                       & 57                                 & 273                                                                                                           & 2180                                                                              \\
codec~\cite{commons-codec}                               & 156                                   & 1780                                  & 992                                                                                    & {\color[HTML]{000000} 91.03}                                                              & 96.01                               & 100                                & 425                                       & 140                                & 680                                                                                                           & 2849                                                                              \\
csv~\cite{commons-csv}                                 & 41                                    & 694                                   & 309                                                                                    & {\color[HTML]{000000} 90.64}                                                              & 92.34                               & 100                                & 146                                       & 35                                 & 235                                                                                                           & 1272                                                                              \\
exec~\cite{commons-exec}                                & 56                                    & 407                                   & 70                                                                                     & {\color[HTML]{000000} 54.84}                                                              & 78.90                               & 100                                & 104                                       & 27                                 & 248                                                                                                           & 327                                                                               \\
fast-pfor~\cite{javafastpfor}                           & 82                                    & 971                                   & 82                                                                                     & {\color[HTML]{000000} 54.55}                                                              & 87.40                               & 100                                & 127                                       & 14                                 & 748                                                                                                           & 302                                                                               \\
fileupload~\cite{commons-fileupload}                          & 49                                    & 381                                   & 39                                                                                     & {\color[HTML]{000000} 13.02}                                                              & 98.31                               & 100                                & 121                                       & 39                                 & 192                                                                                                           & 194                                                                               \\
graph~\cite{commons-graph}                               & 118                                   & 879                                   & 146                                                                                    & {\color[HTML]{000000} 58.78}                                                              & 97.13                               & 100                                & 216                                       & 29                                 & 541                                                                                                           & {\color[HTML]{000000} 975}                                                        \\
jansi~\cite{jansi}                               & 48                                    & 474                                   & 107                                                                                    & {\color[HTML]{000000} 23.47}                                                              & 84.83                               & 100                                & 378                                       & 0                                  & 409                                                                                                           & 123                                                                               \\
pool~\cite{commons-pool}                                & 98                                    & 1097                                  & 73                                                                                     & {\color[HTML]{000000} 22.29}                                                              & 91.60                               & 100                                & 203                                       & 91                                 & 682                                                                                                           & 649                                                                               \\
validator~\cite{commons-validator}                           & 130                                   & 1228                                  & 464                                                                                    & {\color[HTML]{000000} 63.31}                                                              & 95.51                               & 100                                & 421                                       & 209                                & 646                                                                                                           & {\color[HTML]{000000} 1463}                                                       \\ \hline
\textbf{Total}                      & 836                                   & 8575                                  & 2719                                                                                   & {\color[HTML]{000000} 56.57}                                                              & 91.99                               & 100                                & 2245                                      & 641                                & 4654                                                                                                          & {\color[HTML]{000000} 10334}                                                      \\ \hline
\end{tabular}
    \vspace{-10pt}
    \label{table:stats}
\end{table}

Using the universal type mapping, \approach successfully creates and validates project skeletons in the target PL, achieving $100\%$ syntax and runtime validation (column \emph{SV} in Table~\ref{table:stats}). The skeleton validation step ensures all module imports, class structures, method signatures, and type annotations are done properly, making the subsequent steps easier. Applying \approach to unseen projects, if a class skeleton cannot be validated, \approach removes it from the target project, updates the skeleton based on the class dependencies, and proceeds to the next phase. 

\mybox{\textnormal{\textbf{Summary.} \approach can successfully transform projects to remove method and constructor overloading. Moreover, it can automatically translate $91.99\%$ of the source PL types and use that to create and validate project skeletons in the target PL.}}

\subsubsection{Effectiveness in Compositional Translation and Validation:}
\label{subsubsec:rq-effectiveness-translation}

Table~\ref{table:rq1-effectiveness} shows the compositional translation and validation results. The \emph{AMF} column indicates the total number of application method fragments. The numbers in subsequent columns demonstrate the effectiveness of \approach in the translation and validation of \emph{AMF}s only.\footnote{Please refer to our artifact~\cite{website} for details of all the translation and validation of other fragments shown in Table~\ref{table:stats}.} The \textit{Syntax Check} column indicates the percentage of \emph{AMF}s that pass syntactic validation. Column \textit{SNEF} shows the percentage of \emph{AMF}s not covered by source project tests. \approach successfully generates syntactically correct code (\rebuttal{$98.80\%$ of AMFs and $96.40\%$ of all fragments---field, application, and test fragments---across ten subjects}). We also observe that \rebuttal{$43.43\%$} of \emph{AMF}s are not covered during the execution of any test, i.e., we cannot go beyond syntactic check and validate their runtime behavior or functional equivalence. 

For \rebuttal{$56.57\%$} of the \emph{AMF}s that are covered by source project tests, \approach attempts to validate their functional equivalence using GraalVM. Multi-column \textit{GraalVM} shows the percentage of \emph{AMFs} that GraalVM executes and successfully validates (\textit{GS}), executes but there is a test assertion failure (\textit{GF}), and cannot execute due to its limitation (\textit{GE}) mentioned in \S\ref{subsubsec:language-interoperability}. 
\rebuttal{$24.50\%$} (min=\rebuttal{$6.01\%$} and max=\rebuttal{$70.70\%$}), \rebuttal{$16.23\%$} (min=\rebuttal{$1.04\%$} and max=\rebuttal{$32.50\%$}), and \rebuttal{$15.84\%$} (\rebuttal{min=$3.13\%$} and \rebuttal{max=$25.11\%$)} of \emph{AMFs} resulted in \textit{Graal Success}, \textit{Graal Fail}, and \textit{Graal Error}, respectively. Note that these numbers add up to \rebuttal{$56.57\%$} of \emph{AMFs} that were covered by source project tests. With respect to only covered methods, GraalVM Success is \rebuttal{$43.30\%$}. Furthermore, our analysis shows that a high portion of methods that are not covered by tests are either abstract methods or getter/setter methods. If their translations are syntactically correct, they are also likely functionally equivalent, which can ramp up the success rate (\S\ref{sec:rq-augmentation}). Spearman Rank Order Correlation~\cite{spearman1961proof} indicates a strong positive correlation between method coverage (Table~\ref{table:stats}) and \textit{GS} numbers (\rebuttal{$\rho=0.92$}), confirming that with better method coverage, validated \emph{AMF}s are very likely to be higher.

% \reyhan{@Ali, do the correlation analysis and calculate the rho}

\begin{table}
    \setlength{\tabcolsep}{1.3pt}
    \scriptsize
    \centering
    %\vspace{-10pt}
    \caption{Effectiveness of \approach in repository-level code translation. Abbreviations in the table stand for \textbf{AMF}: \#Application Method Fragments, \textbf{SNEF}: Source Non-Exercised Fragments, \textbf{GS}: Graal Success, \textbf{GF}: Graal Fail, \textbf{GE}: Graal Error, \textbf{TNEF}: Target Non-Exercised Fragments, \textbf{ATP}: Fragments All Test Pass, \textbf{OTF}: Fragments One Test Fail, \textbf{MTF}: Fragments Many Test Fail, \textbf{ATF}: Fragments All Test Fail, \textbf{TPR}: Test Pass Rate, \rebuttal{\textbf{O}: Overall}, \textbf{RE}: Runtime Error, \textbf{AF}: Assertion Failure, and \rebuttal{\textbf{M1}: Number of \textit{AMFs} that GraalVM could not execute (\textit{GE}) but translated test fragments exercised.}}
    \vspace{-5pt}
    \begin{tabular}{l|c|c|c|ccc|cccccccccccc|cc}
\hline
                                    &                                &                                                                                          &                                                                                 & \multicolumn{3}{c|}{\textbf{GraalVM}}                                                                        & \multicolumn{12}{c|}{\textbf{Test Translation}}                                                                                                                                                                                                                                                                                                                                                                                                                                                                                                                                                                                                                     & \multicolumn{2}{c}{}                                           \\ \cline{5-19}
                                    &                                &                                                                                          &                                                                                 &                                    &                                    &                                    & \multicolumn{1}{c|}{}                                                                               & \multicolumn{1}{c|}{}                                                                              & \multicolumn{3}{c|}{\textbf{OTF (\%)}}                                                                                 & \multicolumn{3}{c|}{\textbf{MTF (\%)}}                                                                                 & \multicolumn{3}{c|}{\textbf{ATF (\%)}}                                                                                 &                                                                               & \multicolumn{2}{c}{\multirow{-2}{*}{\textbf{M1}}}              \\ \cline{10-18} \cline{20-21} 
\multirow{-3}{*}{\textbf{Subjects}} & \multirow{-3}{*}{\textbf{AMF}} & \multirow{-3}{*}{\textbf{\begin{tabular}[c]{@{}c@{}}Syntax\\ Check\\ (\%)\end{tabular}}} & \multirow{-3}{*}{\textbf{\begin{tabular}[c]{@{}c@{}}SNEF\\  (\%)\end{tabular}}} & \multirow{-2}{*}{\textbf{GS (\%)}} & \multirow{-2}{*}{\textbf{GF (\%)}} & \multirow{-2}{*}{\textbf{GE (\%)}} & \multicolumn{1}{c|}{\multirow{-2}{*}{\textbf{\begin{tabular}[c]{@{}c@{}}TNEF\\ (\%)\end{tabular}}}} & \multicolumn{1}{c|}{\multirow{-2}{*}{\textbf{\begin{tabular}[c]{@{}c@{}}ATP\\ (\%)\end{tabular}}}} & \multicolumn{1}{c|}{\textbf{O}} & \multicolumn{1}{c|}{\textbf{RE}} & \multicolumn{1}{c|}{\textbf{AF}}                  & \multicolumn{1}{c|}{\textbf{O}} & \multicolumn{1}{c|}{\textbf{RE}} & \multicolumn{1}{c|}{\textbf{AF}}                  & \multicolumn{1}{c|}{\textbf{O}} & \multicolumn{1}{c|}{\textbf{RE}} & \multicolumn{1}{c|}{\textbf{AF}}                  & \multirow{-2}{*}{\textbf{\begin{tabular}[c]{@{}c@{}}TPR\\ (\%)\end{tabular}}} & \multicolumn{1}{c|}{\textbf{All}} & \textbf{Some}              \\ \hline
cli                                 & 273                            & {\color[HTML]{000000} 100}                                                               & {\color[HTML]{000000} 5.86}                                                     & {\color[HTML]{000000} 70.70}       & {\color[HTML]{000000} 11.72}       & {\color[HTML]{000000} 11.72}       & \multicolumn{1}{c|}{{\color[HTML]{000000} 35.90}}                                                   & \multicolumn{1}{c|}{{\color[HTML]{000000} 8.42}}                                                   & {\color[HTML]{000000} 10.62}    & {\color[HTML]{000000} 51.72}     & \multicolumn{1}{c|}{{\color[HTML]{000000} 48.28}} & {\color[HTML]{000000} 32.23}    & {\color[HTML]{000000} 91.07}     & \multicolumn{1}{c|}{{\color[HTML]{000000} 8.93}}  & {\color[HTML]{000000} 6.96}     & {\color[HTML]{000000} 100}       & \multicolumn{1}{c|}{{\color[HTML]{000000} 0}}     & {\color[HTML]{000000} 10.08}                                                  & {\color[HTML]{000000} 0}          & {\color[HTML]{000000} 16}  \\
codec                               & 680                            & {\color[HTML]{000000} 98.53}                                                             & {\color[HTML]{000000} 8.97}                                                     & {\color[HTML]{000000} 38.38}       & {\color[HTML]{000000} 32.50}       & {\color[HTML]{000000} 20.15}       & \multicolumn{1}{c|}{{\color[HTML]{000000} 65.59}}                                                   & \multicolumn{1}{c|}{{\color[HTML]{000000} 4.12}}                                                   & {\color[HTML]{000000} 4.41}     & {\color[HTML]{000000} 60.00}     & \multicolumn{1}{c|}{{\color[HTML]{000000} 40.00}} & {\color[HTML]{000000} 12.79}    & {\color[HTML]{000000} 55.11}     & \multicolumn{1}{c|}{{\color[HTML]{000000} 44.89}} & {\color[HTML]{000000} 4.12}     & {\color[HTML]{000000} 75.21}     & \multicolumn{1}{c|}{{\color[HTML]{000000} 24.79}} & {\color[HTML]{000000} 9.43}                                                   & {\color[HTML]{000000} 11}         & {\color[HTML]{000000} 27}  \\
csv                                 & 235                            & {\color[HTML]{000000} 98.72}                                                             & {\color[HTML]{000000} 9.36}                                                     & {\color[HTML]{000000} 38.72}       & {\color[HTML]{000000} 26.81}       & {\color[HTML]{000000} 25.11}       & \multicolumn{1}{c|}{{\color[HTML]{000000} 74.47}}                                                   & \multicolumn{1}{c|}{{\color[HTML]{000000} 0}}                                                      & {\color[HTML]{000000} 13.62}    & {\color[HTML]{000000} 96.88}     & \multicolumn{1}{c|}{{\color[HTML]{000000} 3.13}}  & {\color[HTML]{000000} 0}        & {\color[HTML]{000000} 0}         & \multicolumn{1}{c|}{{\color[HTML]{000000} 0}}     & {\color[HTML]{000000} 2.55}     & {\color[HTML]{000000} 100}       & \multicolumn{1}{c|}{{\color[HTML]{000000} 0}}     & {\color[HTML]{000000} 0}                                                      & {\color[HTML]{000000} 0}          & {\color[HTML]{000000} 3}   \\
exec                                & 248                            & {\color[HTML]{000000} 100}                                                               & {\color[HTML]{000000} 45.16}                                                    & {\color[HTML]{000000} 33.47}       & {\color[HTML]{000000} 2.02}        & {\color[HTML]{000000} 19.35}       & \multicolumn{1}{c|}{{\color[HTML]{000000} 34.27}}                                                   & \multicolumn{1}{c|}{{\color[HTML]{000000} 4.44}}                                                   & {\color[HTML]{000000} 2.02}     & {\color[HTML]{000000} 40.00}     & \multicolumn{1}{c|}{{\color[HTML]{000000} 60.00}} & {\color[HTML]{000000} 7.26}     & {\color[HTML]{000000} 93.36}     & \multicolumn{1}{c|}{{\color[HTML]{000000} 6.64}}  & {\color[HTML]{000000} 6.85}     & {\color[HTML]{000000} 100}       & \multicolumn{1}{c|}{{\color[HTML]{000000} 0}}     & {\color[HTML]{000000} 19.29}                                                  & {\color[HTML]{000000} 6}          & {\color[HTML]{000000} 9}   \\
fast-pfor                           & 748                            & {\color[HTML]{000000} 95.32}                                                             & {\color[HTML]{000000} 45.45}                                                    & {\color[HTML]{000000} 12.03}       & {\color[HTML]{000000} 24.20}       & {\color[HTML]{000000} 18.32}       & \multicolumn{1}{c|}{{\color[HTML]{000000} 41.71}}                                                   & \multicolumn{1}{c|}{{\color[HTML]{000000} 4.28}}                                                   & {\color[HTML]{000000} 1.74}     & {\color[HTML]{000000} 84.62}     & \multicolumn{1}{c|}{{\color[HTML]{000000} 15.38}} & {\color[HTML]{000000} 3.74}     & {\color[HTML]{000000} 79.23}     & \multicolumn{1}{c|}{{\color[HTML]{000000} 20.77}} & {\color[HTML]{000000} 3.07}     & {\color[HTML]{000000} 85.88}     & \multicolumn{1}{c|}{{\color[HTML]{000000} 14.12}} & {\color[HTML]{000000} 20.08}                                                  & {\color[HTML]{000000} 6}          & {\color[HTML]{000000} 25}  \\
fileupload                          & 192                            & {\color[HTML]{000000} 100}                                                               & {\color[HTML]{000000} 86.98}                                                    & {\color[HTML]{000000} 8.85}        & {\color[HTML]{000000} 1.04}        & {\color[HTML]{000000} 3.13}        & \multicolumn{1}{c|}{{\color[HTML]{000000} 1.56}}                                                    & \multicolumn{1}{c|}{{\color[HTML]{000000} 3.65}}                                                   & {\color[HTML]{000000} 6.77}     & {\color[HTML]{000000} 30.77}     & \multicolumn{1}{c|}{{\color[HTML]{000000} 69.23}} & {\color[HTML]{000000} 1.04}     & {\color[HTML]{000000} 91.67}     & \multicolumn{1}{c|}{{\color[HTML]{000000} 8.33}}  & {\color[HTML]{000000} 0}        & {\color[HTML]{000000} 0}         & \multicolumn{1}{c|}{{\color[HTML]{000000} 0}}     & {\color[HTML]{000000} 63.44}                                                  & {\color[HTML]{000000} 2}          & {\color[HTML]{000000} 3}   \\
graph                               & 541                            & {\color[HTML]{000000} 99.63}                                                             & {\color[HTML]{000000} 41.22}                                                    & {\color[HTML]{000000} 24.77}       & {\color[HTML]{000000} 22.92}       & {\color[HTML]{000000} 11.09}       & \multicolumn{1}{c|}{{\color[HTML]{000000} 57.12}}                                                   & \multicolumn{1}{c|}{{\color[HTML]{000000} 0}}                                                      & {\color[HTML]{000000} 0.92}     & {\color[HTML]{000000} 100}       & \multicolumn{1}{c|}{{\color[HTML]{000000} 0}}     & {\color[HTML]{000000} 0.18}     & {\color[HTML]{000000} 100}       & \multicolumn{1}{c|}{{\color[HTML]{000000} 0}}     & {\color[HTML]{000000} 0.55}     & {\color[HTML]{000000} 100}       & \multicolumn{1}{c|}{{\color[HTML]{000000} 0}}     & {\color[HTML]{000000} 11.04}                                                  & {\color[HTML]{000000} 0}          & {\color[HTML]{000000} 1}   \\
jansi                               & 409                            & {\color[HTML]{000000} 99.76}                                                             & {\color[HTML]{000000} 76.53}                                                    & {\color[HTML]{000000} 8.07}        & {\color[HTML]{000000} 11.49}       & {\color[HTML]{000000} 3.91}        & \multicolumn{1}{c|}{{\color[HTML]{000000} 22.25}}                                                   & \multicolumn{1}{c|}{{\color[HTML]{000000} 0.24}}                                                   & {\color[HTML]{000000} 0.98}     & {\color[HTML]{000000} 100}       & \multicolumn{1}{c|}{{\color[HTML]{000000} 0}}     & {\color[HTML]{000000} 0}        & {\color[HTML]{000000} 0}         & \multicolumn{1}{c|}{{\color[HTML]{000000} 0}}     & {\color[HTML]{000000} 0}        & {\color[HTML]{000000} 0}         & \multicolumn{1}{c|}{{\color[HTML]{000000} 0}}     & {\color[HTML]{000000} 1.07}                                                   & {\color[HTML]{000000} 0}          & {\color[HTML]{000000} 1}   \\
pool                                & 682                            & {\color[HTML]{000000} 100}                                                               & {\color[HTML]{000000} 77.71}                                                    & {\color[HTML]{000000} 6.01}        & {\color[HTML]{000000} 1.32}        & {\color[HTML]{000000} 14.96}       & \multicolumn{1}{c|}{{\color[HTML]{000000} 19.35}}                                                   & \multicolumn{1}{c|}{{\color[HTML]{000000} 1.61}}                                                   & {\color[HTML]{000000} 1.03}     & {\color[HTML]{000000} 100}       & \multicolumn{1}{c|}{{\color[HTML]{000000} 0}}     & {\color[HTML]{000000} 0.29}     & {\color[HTML]{000000} 100}       & \multicolumn{1}{c|}{{\color[HTML]{000000} 0}}     & {\color[HTML]{000000} 0}        & {\color[HTML]{000000} 0}         & \multicolumn{1}{c|}{{\color[HTML]{000000} 0}}     & {\color[HTML]{000000} 6.62}                                                   & {\color[HTML]{000000} 4}          & {\color[HTML]{000000} 2}   \\
validator                           & 646                            & {\color[HTML]{000000} 99.23}                                                             & {\color[HTML]{000000} 36.69}                                                    & {\color[HTML]{000000} 30.50}       & {\color[HTML]{000000} 11.15}       & {\color[HTML]{000000} 21.67}       & \multicolumn{1}{c|}{{\color[HTML]{000000} 46.44}}                                                   & \multicolumn{1}{c|}{{\color[HTML]{000000} 3.25}}                                                   & {\color[HTML]{000000} 5.42}     & {\color[HTML]{000000} 71.43}     & \multicolumn{1}{c|}{{\color[HTML]{000000} 28.57}} & {\color[HTML]{000000} 4.18}     & {\color[HTML]{000000} 84.50}     & \multicolumn{1}{c|}{{\color[HTML]{000000} 15.50}} & {\color[HTML]{000000} 4.02}     & {\color[HTML]{000000} 99.12}     & \multicolumn{1}{c|}{{\color[HTML]{000000} 0.88}}  & {\color[HTML]{000000} 11.70}                                                  & {\color[HTML]{000000} 1}          & {\color[HTML]{000000} 31}  \\ \hline
\textbf{Total}                      & 4654                           & {\color[HTML]{000000} 98.80}                                                             & {\color[HTML]{000000} 43.43}                                                    & {\color[HTML]{000000} 24.50}       & {\color[HTML]{000000} 16.23}       & {\color[HTML]{000000} 15.84}       & \multicolumn{1}{c|}{{\color[HTML]{000000} 41.92}}                                                   & \multicolumn{1}{c|}{{\color[HTML]{000000} 2.88}}                                                   & {\color[HTML]{000000} 3.72}     & {\color[HTML]{000000} 70.52}     & \multicolumn{1}{c|}{{\color[HTML]{000000} 29.48}} & {\color[HTML]{000000} 5.44}     & {\color[HTML]{000000} 82.77}     & \multicolumn{1}{c|}{{\color[HTML]{000000} 17.23}} & {\color[HTML]{000000} 2.62}     & {\color[HTML]{000000} 92.86}     & \multicolumn{1}{c|}{{\color[HTML]{000000} 7.14}}  & {\color[HTML]{000000} 9.76}                                                   & {\color[HTML]{000000} 30}         & {\color[HTML]{000000} 118} \\ \hline
\end{tabular}
    \label{table:rq1-effectiveness}
    \vspace{-10pt}
\end{table}

Regardless of GraalVM's validation for functional equivalence, \approach translates and executes the test fragments on the recomposed translated project. The \emph{Test Translation} multi-column shows the results of test translation and execution. Column \textit{TNEF} indicates the ratio of \emph{AMF}s where execution of translated tests never reached them. Columns \textit{ATP} through \textit{ATF} show the number of \emph{AMFs} that \approach executed using translated test fragments, categorized per the test execution results. For \rebuttal{$2.88\%$} of the \emph{AMFs}, all the test fragments that covered them were marked as pass (\textit{ATP}). For \rebuttal{$3.72\%$} and \rebuttal{$5.44\%$}, at least one test (\textit{OTF}) or more than one test (\textit{MTF}) failed. All the test fragments failed for \rebuttal{$2.62\%$} of the \emph{AMFs}. \rebuttal{Note that these numbers (TNEF, ATP, OTF, MTF, and ATF) add up to $56.57\%$ of the AMFs that are covered by source project tests.}

For cases with test failure, columns \emph{RE} and \emph{AF} show the breakdown of whether test failure was due to assertion failure or runtime error. We can observe that most test executions terminated with a runtime error due to translation bugs and never reached an assert statement. \emph{Our manual investigation confirms that a high rate of runtime errors is due to a relatively small number of fragments with translation bugs.} Although test decomposition helps with \emph{test translation coupling effect} (\S\ref{sec:illustrative-example}), there is still a high degree of runtime errors due to long call chains in these projects (the average number of methods executed per test in the original and decomposed test suites are $27.4$ and $21.8$, respectively). As a result, the overall pass rate, i.e., the percentage of recomposed test fragments for the translated projects that pass (column \emph{TPR}), is low. \rebuttal{These results also confirm the necessity of in-isolation testing through language interoperability by \approach.}

We also calculated the number of \emph{AMFs} that GraalVM could not execute (numbers under \textit{GE} column) but translated test fragments exercised (column \emph{M1}). \textit{All} indicates the number of \emph{AMF}s with all passing tests, including test fragments with assert statements, indicating the validation of functional equivalence (with respect to the source project tests). \textit{Some} corresponds to the number of \emph{AMF}s with at least one passing test, which indicates runtime validation. Overall, test translation validates the functional correctness and runtime behavior of \rebuttal{$30$} and \rebuttal{$118$} fragments that GraalVM could not exercise. 

\rebuttal{Finally, we analyzed translations using PyLint~\cite{pylint}, which scores Python files on a scale of $0$ to $10$ based on how Pythonic the code is.
%, following PEP8 rules~\cite{pep8}. 
All \approach translations achieved scores of $10$, mainly because (1) LLMs inherently generate idiomatic code and (2) AlphaTrans uses Black~\cite{blakc} for formatting the translations extracted from the LLM response. These results confirm that the translations are all Pythonic, i.e., they adhere to Python coding standards and best practices.}

% Total AMF = 4654
% GS = 1140
% M1 All = 30
% M1 Some = 118
% runtime validation = (GS + M1 Some) / Total AMF = (1140 + 118) / 4654 = 27.03 %
% functional equivalence = (GS + M1 All) / Total AMF = (1140 + 30) / 4654 = 25.14%

\vspace{-3pt}
\mybox{\textnormal{\textbf{Summary.} \approach effectively performs compositional translation and validation of \rebuttal{$17,874$ fragments}, achieving overall \rebuttal{$96.40\%$ syntactic correctness ($98.80\%$ for AMFs)}, \rebuttal{$27.03\%$ runtime behavior validation (GS+M1 Some)}, and \rebuttal{$25.14\%$ functional equivalence (GS+M1 All)}.}}
\vspace{-5pt}

\subsection{RQ2: Translation Bugs and Fixes}
\label{sec:rq-bugs}

We investigated the manual effort for fixing translation bugs in a subset of studied subjects, namely \textit{Commons-FileUpload}, \textit{Commons-CLI}, \textit{Commons-CSV}, and \textit{Commons-Validator}. We also discuss some of the translation bugs and fixes for them to better illustrate the challenges in code translation.  

\vspace{-5pt}
\subsubsection{Human Study}
Our two human subjects were selected due to their relative familiarity with the selected projects. Their effort indicates an upper bound for the amount of time required to fix translation bugs since developers of the source projects are likely to fix the bugs better and faster. We shared with them the source program in Java, the translations by \approach, and all the reports and artifacts generated by \approach during translation. 

For \textit{Commons-FileUpload}, achieving green tests took $5.5$ hours and required $120$ and $114$ line additions and deletions from partial translations. For \textit{Commons-CLI}, the manual fix took $11$ hours, making $614$ and $1,253$ line additions and deletions, respectively. The project was very dense for \textit{Commons-CSV}, with many method calls, making it harder to manually fix bugs. Nevertheless, a developer achieved all green tests in $30$ hours with $2,676$ and $999$ line additions and deletions, respectively. Finally, for \textit{Commons-Validator}, the developer spent $34$ hours to fix translation bugs, with $3,585$ and $2,416$ line additions and deletions, respectively. One of the major feedback from developers was that test decomposition greatly helped locate and fix translation bugs: in case of a test failure, developers only need to investigate the last call statement in the failed test fragment instead of looking at the stack trace and other prior calls.
%(more quantified details about the impact of test decomposition on validation in \S\ref{sec:rq-test-decomposition}). 

\vspace{-5pt}
\subsubsection{Translation Bugs}
Our artifacts~\cite{website} contain partial translations and fixed versions as separate commits. These commits can serve as useful benchmarks for evaluating fault localization, program repair, and test generation techniques. This section shows \emph{four} instances of such translation bugs. 
%a subset of translation bugs and the fix provided by the developers, which can be helpful for future research to advance code translation. 
%In this section, we show some of interesting examples of bugs that we identified in two different aspects of repository-level code translation and discuss how a human developer would fix them, which can lead to building a more advanced automated approach in the future.
%\begin{itemize}[leftmargin=*]
 %\item 
 %\textit{Method translation bugs.} 
 
 The two most prevalent sources of translation bugs are mismatches between APIs and behavioral differences in the PLs. The code snippet below demonstrates a bug that happened due to a mismatch in the logic of \texttt{\small Calendar} (Java) and \texttt{\small datetime} (Python). Line $3$ in Java sets the \texttt{\small MONTH} field to \texttt{\small 0}, which corresponds to the first month of the year (January). Similarly, the Python translation sets the \texttt{\small month} attribute to \texttt{\small 0}; however, in the Python library, January corresponds to value~1 for \texttt{\small month}. % the first month, i.e., the correct translation should use index \texttt{\small 1}.

\vspace{3pt}
\noindent
\begin{minipage}{.475\linewidth}
\begin{minted}[frame=lines,framesep=1mm,baselinestretch=0.5, fontsize=\scriptsize, breaklines, breakanywhere, linenos,numbersep=2pt]{java}
        ----------- JAVA SOURCE CODE -----------
Calendar calendar = Calendar.getInstance()
calendar.set(Calendar.MONTH, 0);
\end{minted}
% \end{lstlisting}
\end{minipage}\hfill
\begin{minipage}{.475\linewidth}
\begin{minted}[escapeinside=||, frame=lines,framesep=1mm,baselinestretch=0.5, fontsize=\scriptsize, breaklines, breakanywhere, linenos,numbersep=2pt, highlightlines={3,4}]{py}
        ---------- PYTHON TRANSLATION ----------
  calendar = datetime.datetime.now()
|\xglobal\colorlet{FancyVerbHighlightColor}{bubblegum}|- calendar = calendar.replace(month=0)
|\xglobal\colorlet{FancyVerbHighlightColor}{cambridgeblue}|+ calendar = calendar.replace(month=1)
\end{minted}
% \end{lstlisting}
\end{minipage}
\vspace{3pt}

The next example shows the difference in implicit type casting between the two Pls. Line $5$ in Java source code concatenates a \texttt{\small String} value with \texttt{\small null}. During execution, Java runtime silently casts \texttt{\small null} to a \texttt{\small String} and then performs the concatenation operation on it. In Python, concatenating an \texttt{\small str} with \texttt{\small None} results in a \texttt{\small TypeError} as the operands of the binary operation has different types. A correct Python translation requires explicit casting of \texttt{\small None} to \texttt{\small str} as shown in Line $5$.

\vspace{5pt}
\noindent
\begin{minipage}{.475\linewidth}
% \begin{lstlisting}[language = source, columns=fullflexible, basicstyle=\scriptsize\ttfamily,showlines=true]
\begin{minted}[frame=lines,framesep=1mm,baselinestretch=0.5, fontsize=\scriptsize, breaklines, breakanywhere, linenos,numbersep=2pt]{java}
        ----------- JAVA SOURCE CODE -----------
qChar = "'";
nullStr = null;

this.qNullStr = qChar + nullStr + qChar;
\end{minted}
% \end{lstlisting}
\end{minipage}\hfill
\begin{minipage}{.475\linewidth}
% \begin{lstlisting}[language = source, columns=fullflexible, basicstyle=\scriptsize\ttfamily,showlines=true]
\begin{minted}[escapeinside=||, frame=lines,framesep=1mm,baselinestretch=0.5, fontsize=\scriptsize, breaklines, breakanywhere, linenos,numbersep=2pt, highlightlines={4,5}]{py}
        ---------- PYTHON TRANSLATION ----------
  qChar = "'"
  nullStr = None
|\xglobal\colorlet{FancyVerbHighlightColor}{bubblegum}|- self.qNullStr = qChar + nullStr + qChar
|\xglobal\colorlet{FancyVerbHighlightColor}{cambridgeblue}|+ self.qNullStr = qChar + str(nullStr) + qChar
\end{minted}
% \end{lstlisting}
\end{minipage}
\vspace{5pt}

The third example shows an instance of \texttt{\small write(int b)} method from \texttt{\small ByteArrayOutputStream} class, where the least significant $8$ bits of the integer \texttt{\small (b2 << 4) | (b3 >> 2)} are directly written to the stream. The incorrect Python translation attempts to construct a \texttt{\small bytes} object using a singleton list with the input integer before writing it to an object of type \texttt{\small io.BytesIO}. However, this neglects that the \texttt{\small bytes()} constructor requires the integers in the input iterable to be strictly in the range of \texttt{\small [0, 255]}. Thereby, a \texttt{\small ValueError} is thrown when \texttt{\small b2} is large. The correct translation requires \texttt{\small 0xF}, a mask that maintains only the 4 lowest bits of \texttt{\small b2} before left-shifting by 4 as shown in Line $3$ under Python translation. Given that \texttt{\small b3} and \texttt{\small b4} each contain no more than 8 bits, this change ensures the least significant $8$ bits of \texttt{\small (b2 << 4) | (b3 >> 2)} are correctly written to the \texttt{\small BytesIO} object.

\vspace{5pt}
\noindent
\begin{minipage}{.475\linewidth}
% \begin{lstlisting}[language = source, columns=fullflexible, basicstyle=\scriptsize\ttfamily,showlines=true]
\begin{minted}[frame=lines,framesep=1mm,baselinestretch=0.5, fontsize=\scriptsize, breaklines, breakanywhere, linenos,numbersep=2pt]{java}
        ----------- JAVA SOURCE CODE -----------

out.write((b2 << 4) | (b3 >> 2));
\end{minted}
% \end{lstlisting}
\end{minipage}\hfill
\begin{minipage}{.475\linewidth}
% \begin{lstlisting}[language = source, columns=fullflexible, basicstyle=\scriptsize\ttfamily,showlines=true]
\begin{minted}[escapeinside=@@, frame=lines,framesep=1mm,baselinestretch=0.5, fontsize=\scriptsize, breaklines, breakanywhere, linenos,numbersep=2pt, highlightlines={2,3}]{py}
        ---------- PYTHON TRANSLATION ----------
@\xglobal\colorlet{FancyVerbHighlightColor}{bubblegum}@- out.write(bytes([(b2 << 4) | (b3 >> 2)]))
@\xglobal\colorlet{FancyVerbHighlightColor}{cambridgeblue}@+ out.write(bytes([((b2 & 0xF) << 4) | (b3 >> 2)]))
\end{minted}
% \end{lstlisting}
\end{minipage}
\vspace{5pt}

The last code snippet demonstrates the Java behavior of an iterator that is unavailable in Python. The incorrect Python translation uses \texttt{\small next()} to implement both \texttt{\small next()} and \texttt{\small hasNext()} methods of the Java \texttt{\small java.util.Iterator} type. However, calling \texttt{\small next()} increments the iterator in Python. The correct translation should implement \texttt{\small PeekableIterator} interface in Python with a method  \texttt{\small hasNext() -> bool}.

\vspace{5pt}
\noindent
\begin{minipage}{.475\linewidth}
% \begin{lstlisting}[language = source, columns=fullflexible, basicstyle=\scriptsize\ttfamily,showlines=true]
\begin{minted}[frame=lines,framesep=1mm,baselinestretch=0.5, fontsize=\scriptsize, breaklines, breakanywhere, linenos,numbersep=2pt]{java}
        ----------- JAVA SOURCE CODE -----------
Iterator<String> headers = ls.keySet().iterator();

assertEquals("content", headers.next());

assertFalse(headers.hasNext());
\end{minted}
% \end{lstlisting}
\end{minipage}\hfill
\begin{minipage}{.475\linewidth}
% \begin{lstlisting}[language = source, columns=fullflexible, basicstyle=\scriptsize\ttfamily,showlines=true]
\begin{minted}[escapeinside=@@, frame=lines,framesep=1mm,baselinestretch=0.5, fontsize=\scriptsize, breaklines, breakanywhere, linenos,numbersep=2pt, highlightlines={2,3, 5, 6}]{py}
        ---------- PYTHON TRANSLATION ----------
@\xglobal\colorlet{FancyVerbHighlightColor}{bubblegum}@- headers = iter(ls.keys())
@\xglobal\colorlet{FancyVerbHighlightColor}{cambridgeblue}@+ headers = PeekableIterator(ls.keys())
  self.assertEqual("content", next(headers))
@\xglobal\colorlet{FancyVerbHighlightColor}{bubblegum}@- self.assertFalse(next(headers, None) is not None)
@\xglobal\colorlet{FancyVerbHighlightColor}{cambridgeblue}@+ self.assertFalse(headers.hasNext())
\end{minted}
% \end{lstlisting}
\end{minipage}
\vspace{5pt}

\textbf{Implications.} To obtain correct translation, especially for code using library APIs, models need to generate test cases as well that can validate the translated fragment in isolation. This could be an interesting direction for applying an agentic approach, where the orchestrating agent can decide when to generate test cases, and the test case generator agent gets all the information by running static analysis tools, gathering context from previous runs, collecting API documentation by crawling the internet, and finally generating the translation based on all the information. 

\vspace{-5pt}
\mybox{\textnormal{\textbf{Summary.} Although \approach cannot validate all the translations, it provides partial translations and artifacts that developers can use to complete the translation and achieve green tests in a reasonable time (20.1 hours, on average).}}
\vspace{-10pt}

\subsection{RQ3: Impact of Test Decomposition}
\label{sec:rq-test-decomposition}

\begin{wrapfigure}[12]{r}{0.5\textwidth}
    \centering
    \vspace{-5pt}
    \includegraphics[width=\linewidth]{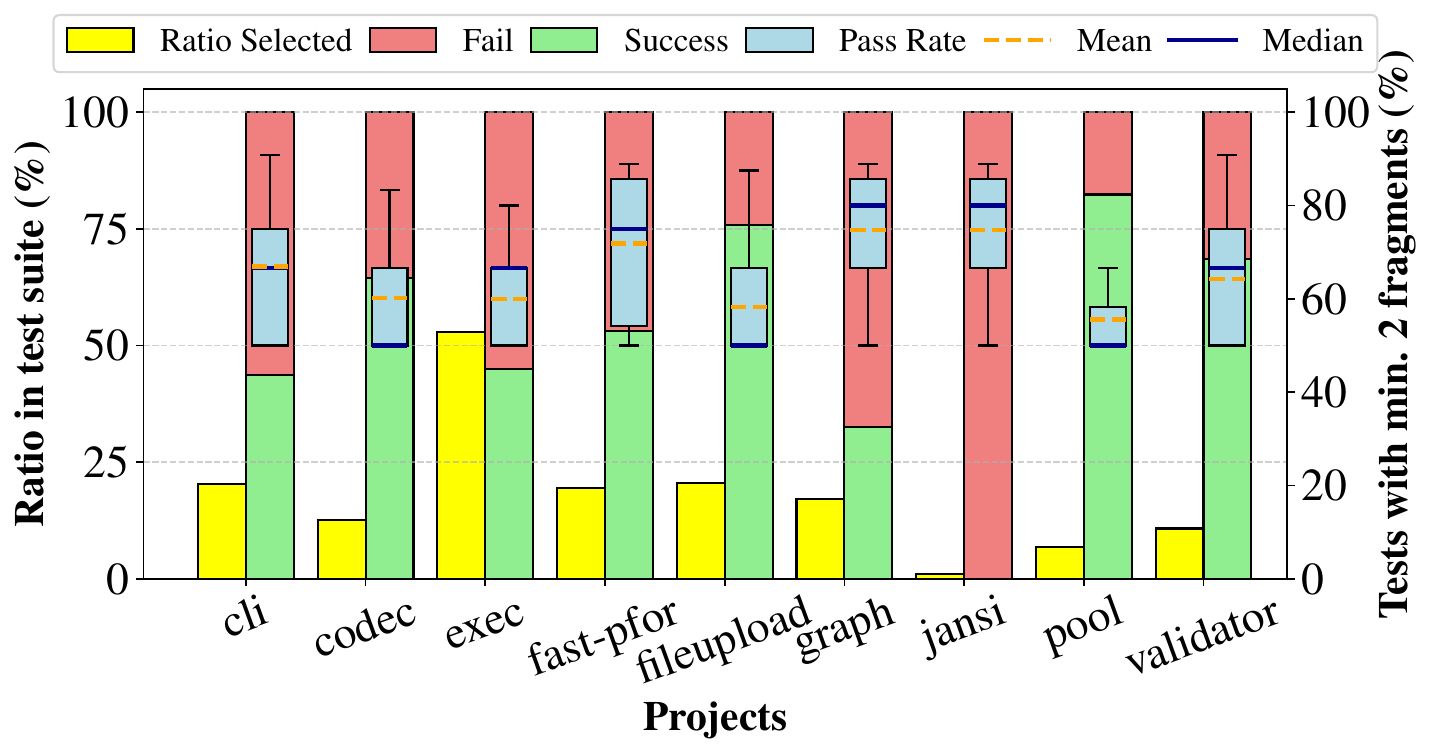}
    \vspace{-18pt}
    \caption{Effectiveness of test decomposition in \approach for validating earlier fragments in failing tests.}
    \label{fig:test-decomposition}
    \vspace{-5pt} % to add the whitespace below
\end{wrapfigure}

We previously showed the effectiveness of test translation in validating the runtime behavior or functional correctness of translated method fragments (\S\ref{subsubsec:rq-effectiveness-translation}). 
To better understand how test decomposition helps address \emph{test translation coupling effect}, we collected translated unit tests with at least two decomposed test fragments. We further applied filtering and kept only those tests for which \emph{all} their decomposed fragments were executed, regardless of passing or failing outcomes. The yellow bars in Figure~\ref{fig:test-decomposition} show the percentage of selected unit tests from the translated test suites.  
None of the tests met the criteria for \textit{Commons-CSV}, so we excluded \rebuttal{it} from further investigation.

We categorized the selected unit tests into two groups: those with all passing test fragments (green bars in Figure~\ref{fig:test-decomposition}) and those with at least one failing test fragment (red bars in Figure~\ref{fig:test-decomposition}). For the unit tests in the latter group, we calculated the pass rate of the \emph{decomposed test fragments}. The blue box chart in Figure~\ref{fig:test-decomposition} shows the distribution of the measured pass rate per unit test. These unit tests would have been marked as \emph{fail} without test decomposition. However, we can observe that these tests can be decomposed into test fragments such that \rebuttal{$62.41\%$} of them pass. These results indicate how decomposed test fragments were useful in helping developers localize translation bugs more easily and resolve translation bugs faster.

\mybox{\textnormal{\textbf{Summary.} Test decomposition unburdens validation of translated method fragments from incorrect translations. \rebuttal{$62.41\%$} of test fragments for unit tests that would have been marked as failed achieved passing outcomes.}}

%\vspace{-15pt}
\subsection{RQ4: Impact of Test Coverage}
\label{sec:rq-augmentation}

Although existing developer-written tests are useful for checking functional equivalence, they can pose two major issues for automated code translation and validation. 
%Our results from RQ1 suggest that existing developer-written tests indicate two major problems. 
First, the coverage for these tests can be extremely low (e.g., \rebuttal{$13.02\%$ for Commons-FileUpload~\cite{commons-fileupload} as reported in Table~\ref{table:stats}}), preventing most of the code from being validated; as our investigation of RQ1 showed, the translation validation rate strongly correlates with test suites' (method) coverage. Second, a developer-written test can have a long call sequence. To show the positive impact of more-focused tests with higher coverage on translation validation, we automatically generated additional tests using EvoSuite~\cite{fraser2011evosuite}. We used default tool configuration %DynaMOSA~\cite{panichella2018dynamosa} as an optimization algorithm
with a time budget of $120$ seconds per class. 

\begin{table}[t]
    \setlength{\tabcolsep}{1pt}
    \scriptsize
    \centering
    \caption{Effectiveness of test augmentation in exercising and validating more application method fragments. \rebuttal{Abbreviations in the table stand for \textbf{ATP}: Fragments All Test Pass and  \textbf{TPR}: Test Pass Rate. \rebuttal{ATP+ and TPR+ demonstrate ATP and TPR gain through test augmentation.}}}
    \vspace{-7pt}
    \begin{tabular}{l|ccccc|ccccc}
\hline
                                    & \multicolumn{5}{c|}{\textbf{Developer-Written Test}}                                                                                                                                                                                                                                                                                                                                  & \multicolumn{5}{c}{\textbf{EvoSuite Test}}                                                                                                                                                                                                                                                                                                                \\ \cline{2-11} 
\multirow{-2}{*}{\textbf{Subjects}} & \textbf{\begin{tabular}[c]{@{}c@{}}Method\\ Coverage (\%)\end{tabular}} & \textbf{\begin{tabular}[c]{@{}c@{}}\# Decomposed\\ Tests\end{tabular}} & \textbf{\begin{tabular}[c]{@{}c@{}}Avg. Methods\\ Executed / Test\end{tabular}} & \textbf{\begin{tabular}[c]{@{}c@{}}TPR\\ (\%)\end{tabular}} & {\color[HTML]{000000} \textbf{\begin{tabular}[c]{@{}c@{}}ATP\\ (\%)\end{tabular}}} & \textbf{\begin{tabular}[c]{@{}c@{}}Method\\ Coverage (\%)\end{tabular}} & \textbf{\# Tests} & \textbf{\begin{tabular}[c]{@{}c@{}}Avg. Methods\\ Executed / Test\end{tabular}} & {\color[HTML]{000000} \textbf{\begin{tabular}[c]{@{}c@{}}TPR+\\ (\%)\end{tabular}}} & {\color[HTML]{000000} \textbf{\begin{tabular}[c]{@{}c@{}}ATP+\\ (\%)\end{tabular}}} \\ \hline
cli                                 & {\color[HTML]{000000} 94.14}                                            & 3036                                                                   & 34.25                                                                           & {\color[HTML]{000000} 10.08}                                & {\color[HTML]{000000} 8.42}                                                        & {\color[HTML]{000000} 95.97}                                            & 569               & 12.15                                                                           & {\color[HTML]{000000} 2.99}                                                         & {\color[HTML]{000000} 1.47}                                                         \\
codec                               & {\color[HTML]{000000} 91.03}                                            & 3522                                                                   & 10.56                                                                           & {\color[HTML]{000000} 9.43}                                 & {\color[HTML]{000000} 4.12}                                                        & {\color[HTML]{000000} 80.74}                                            & 1141              & 8.02                                                                            & {\color[HTML]{000000} 3.51}                                                         & {\color[HTML]{000000} 0.88}                                                         \\
csv                                 & {\color[HTML]{000000} 90.64}                                            & 1219                                                                   & 52.62                                                                           & {\color[HTML]{000000} 0}                                    & {\color[HTML]{000000} 0}                                                           & {\color[HTML]{000000} 74.04}                                            & 220               & 39.16                                                                           & {\color[HTML]{000000} 0}                                                            & {\color[HTML]{000000} 0.00}                                                         \\
exec                                & {\color[HTML]{000000} 54.84}                                            & 311                                                                    & 18.99                                                                           & {\color[HTML]{000000} 19.29}                                & {\color[HTML]{000000} 4.44}                                                        & {\color[HTML]{000000} 61.29}                                            & 245               & 6.32                                                                            & {\color[HTML]{000000} 3.27}                                                         & {\color[HTML]{000000} 1.21}                                                         \\
fast-pfor                           & {\color[HTML]{000000} 54.55}                                            & 249                                                                    & 41.62                                                                           & {\color[HTML]{000000} 20.08}                                & {\color[HTML]{000000} 4.28}                                                        & {\color[HTML]{000000} 39.17}                                            & 1843              & 4.31                                                                            & {\color[HTML]{000000} 5.59}                                                         & {\color[HTML]{000000} 1.07}                                                         \\
fileupload                          & {\color[HTML]{000000} 13.02}                                            & 93                                                                     & 3.54                                                                            & {\color[HTML]{000000} 63.44}                                & {\color[HTML]{000000} 3.65}                                                        & {\color[HTML]{000000} 70.31}                                            & 231               & 5.29                                                                            & {\color[HTML]{000000} 11.26}                                                        & {\color[HTML]{000000} 2.60}                                                         \\
graph                               & {\color[HTML]{000000} 58.78}                                            & 933                                                                    & 25.02                                                                           & {\color[HTML]{000000} 11.04}                                & {\color[HTML]{000000} 0.00}                                                        & {\color[HTML]{000000} 76.71}                                            & 800               & 9.00                                                                            & {\color[HTML]{000000} 5.13}                                                         & {\color[HTML]{000000} 0.92}                                                         \\
jansi                               & {\color[HTML]{000000} 23.47}                                            & 187                                                                    & 13.57                                                                           & {\color[HTML]{000000} 1.07}                                 & {\color[HTML]{000000} 0.24}                                                        & {\color[HTML]{000000} 51.83}                                            & 332               & 9.08                                                                            & {\color[HTML]{000000} 3.31}                                                         & {\color[HTML]{000000} 0.73}                                                         \\
pool                                & {\color[HTML]{000000} 22.29}                                            & 287                                                                    & 6.52                                                                            & {\color[HTML]{000000} 6.62}                                 & {\color[HTML]{000000} 1.61}                                                        & {\color[HTML]{000000} 37.24}                                            & 394               & 7.36                                                                            & {\color[HTML]{000000} 10.41}                                                        & {\color[HTML]{000000} 2.20}                                                         \\
validator                           & {\color[HTML]{000000} 63.31}                                            & {\color[HTML]{000000} 1479}                                            & 11.68                                                                           & {\color[HTML]{000000} 11.70}                                & {\color[HTML]{000000} 3.25}                                                        & {\color[HTML]{000000} 81.42}                                            & 1305              & 13.43                                                                           & {\color[HTML]{000000} 9.73}                                                         & {\color[HTML]{000000} 7.59}                                                         \\ \hline
\textbf{Total}                      & {\color[HTML]{000000} 56.57}                                            & {\color[HTML]{000000} 11316}                                           & 21.84                                                                           & {\color[HTML]{000000} 9.76}                                 & {\color[HTML]{000000} 2.88}                                                        & {\color[HTML]{000000} 66.87}                                            & 7080              & 11.41                                                                           & {\color[HTML]{000000} 5.85}                                                         & {\color[HTML]{000000} 2.11}                                                         \\ \hline
\end{tabular}
    \label{table:rq4-augmentation}
    \vspace{-5pt}
\end{table}

Table~\ref{table:rq4-augmentation} compares properties of the developer-written and EvoSuite-generated tests. Evosuite tests cover more methods than the developer-written test suite (\rebuttal{$66.87\%$} \emph{Method Coverage} compared to \rebuttal{$56.57\%$}). 
%Since we configure EvoSuite to generate several tests per each Java class, the average \emph{Method Coverage} of its tests is higher than developer-written tests (\rebuttal{$66.9\%$} compared to \rebuttal{$56.6\%$}). 
Furthermore, the average number of methods executed per single test is almost half that of decomposed test suites ($11.41$ compared to $21.84$ methods). \rebuttal{To demonstrate the impact of test quality on \approach's overall performance, we translated and executed the EvoSuite-generated tests. Corroborated by the numbers under \emph{TPR+} and \emph{ATP+} columns, we can see that augmenting the test suite increases the method coverage, and thereby, the TPR and ATP numbers from RQ1 by $5.85\%$ and $2.11\%$, respectively.}
%we can see that \approach can further validate the correctness of $2.3\%$ of fragments not exercised by developer tests, increasing the overall functional equivalence to \hl{$XX\%$}. 
%Using test augmentation also achieves an extra $10.2\%$ test pass rate.} 
Not all EvoSuite tests have assertions, and even if they do, the quality of the assertions could be lower compared to developer-written tests (e.g., checking trivial properties with weak fault detection ability). Nonetheless, the higher \emph{TPR} and \emph{ATP} of such tests enhance runtime validation, which is still promising in code translation. EvoSuite is incompatible with Java $21$, and hence GraalVM, which prevented us from using Evosuite-generated tests in \approach. We anticipate that incorporating it in \approach could improve the overall quality of translations. 

%\vspace{-5pt}
\mybox{\textnormal{\textbf{Summary.} %\rebuttal{Augmenting the existing test suite can further validate the correctness of $2.3\%$ of fragments not executed by developer tests.}}}
\rebuttal{Augmenting the existing test suite increases code coverage, thereby exercising and validating more AMFs. Test augmentation can further validate the correctness of $2.11\%$ of fragments not executed by developer tests. The generated tests are more focused and, on average, invoke $48\%$ fewer methods than the developer-written tests.}}}

\subsection{RQ5: Ablation Study}
\label{sec:rq-ablation}

\begin{table}[t]
    \centering
    \scriptsize
    \caption{\rebuttal{Importance of program transformation. Abbreviations in the table stand for \textbf{AMF}: \#Application Method Fragments, \textbf{SNEF}: Source Non-Exercised Fragments, \textbf{GS}: Graal Success, \textbf{GF}: Graal Fail, \textbf{GE}: Graal Error, \textbf{TNEF}: Target Non-Exercised Fragments, \textbf{ATP}: Fragments All Test Pass, \textbf{OTF}: Fragments One Test Fail, \textbf{MTF}: Fragments Many Test Fail, \textbf{ATF}: Fragments All Test Fail, and \textbf{TPR}: Test Pass Rate.}}
    % \reyhan{Ali, please change the color of text in this table to blue}
    \vspace{-7pt}
    \begin{tabular}{l|c|c|ccc|cccccc}
\hline
{\color[HTML]{000000} }                                    & {\color[HTML]{000000} }                               & {\color[HTML]{000000} }                                                                                & \multicolumn{3}{c|}{{\color[HTML]{000000} \textbf{GraalVM}}}                                                                                                                                                                                                 & \multicolumn{6}{c}{{\color[HTML]{000000} \textbf{Test Translation}}}                                                                                                                                                                                                                                                                                                                                                                                                                                                                                                                                                                     \\ \cline{4-12} 
\multirow{-2}{*}{{\color[HTML]{000000} \textbf{Subjects}}} & \multirow{-2}{*}{{\color[HTML]{000000} \textbf{AMF}}} & \multirow{-2}{*}{{\color[HTML]{000000} \textbf{\begin{tabular}[c]{@{}c@{}}SNEF\\  (\%)\end{tabular}}}} & {\color[HTML]{000000} \textbf{\begin{tabular}[c]{@{}c@{}}GS \\ (\%)\end{tabular}}} & {\color[HTML]{000000} \textbf{\begin{tabular}[c]{@{}c@{}}GF \\ (\%)\end{tabular}}} & {\color[HTML]{000000} \textbf{\begin{tabular}[c]{@{}c@{}}GE \\ (\%)\end{tabular}}} & \multicolumn{1}{c|}{{\color[HTML]{000000} \textbf{\begin{tabular}[c]{@{}c@{}}TNEF\\ (\%)\end{tabular}}}} & \multicolumn{1}{c|}{{\color[HTML]{000000} \textbf{\begin{tabular}[c]{@{}c@{}}ATP\\ (\%)\end{tabular}}}} & \multicolumn{1}{c|}{{\color[HTML]{000000} \textbf{\begin{tabular}[c]{@{}c@{}}OTF \\ (\%)\end{tabular}}}} & \multicolumn{1}{c|}{{\color[HTML]{000000} \textbf{\begin{tabular}[c]{@{}c@{}}MTF \\ (\%)\end{tabular}}}} & \multicolumn{1}{c|}{{\color[HTML]{000000} \textbf{\begin{tabular}[c]{@{}c@{}}ATF \\ (\%)\end{tabular}}}} & {\color[HTML]{000000} \textbf{\begin{tabular}[c]{@{}c@{}}TPR\\ (\%)\end{tabular}}} \\ \hline
{\color[HTML]{000000} cli}                                 & {\color[HTML]{000000} 276}                            & {\color[HTML]{000000} 5.80}                                                                            & {\color[HTML]{000000} 0}                                                           & {\color[HTML]{000000} 0}                                                           & {\color[HTML]{000000} 94.20}                                                       & \multicolumn{1}{c|}{{\color[HTML]{000000} 85.51}}                                                        & \multicolumn{1}{c|}{{\color[HTML]{000000} 0}}                                                           & \multicolumn{1}{c|}{{\color[HTML]{000000} 2.17}}                                                         & \multicolumn{1}{c|}{{\color[HTML]{000000} 1.09}}                                                         & \multicolumn{1}{c|}{{\color[HTML]{000000} 5.43}}                                                         & {\color[HTML]{000000} 0.46}                                                        \\
{\color[HTML]{000000} codec}                               & {\color[HTML]{000000} 678}                            & {\color[HTML]{000000} 10.62}                                                                           & {\color[HTML]{000000} 0}                                                           & {\color[HTML]{000000} 0}                                                           & {\color[HTML]{000000} 89.38}                                                       & \multicolumn{1}{c|}{{\color[HTML]{000000} 72.86}}                                                        & \multicolumn{1}{c|}{{\color[HTML]{000000} 2.06}}                                                        & \multicolumn{1}{c|}{{\color[HTML]{000000} 5.46}}                                                         & \multicolumn{1}{c|}{{\color[HTML]{000000} 2.80}}                                                         & \multicolumn{1}{c|}{{\color[HTML]{000000} 6.19}}                                                         & {\color[HTML]{000000} 5.04}                                                        \\
{\color[HTML]{000000} csv}                                 & {\color[HTML]{000000} 235}                            & {\color[HTML]{000000} 8.51}                                                                            & {\color[HTML]{000000} 0}                                                           & {\color[HTML]{000000} 0}                                                           & {\color[HTML]{000000} 91.49}                                                       & \multicolumn{1}{c|}{{\color[HTML]{000000} 88.09}}                                                        & \multicolumn{1}{c|}{{\color[HTML]{000000} 0.43}}                                                        & \multicolumn{1}{c|}{{\color[HTML]{000000} 0.43}}                                                         & \multicolumn{1}{c|}{{\color[HTML]{000000} 0.85}}                                                         & \multicolumn{1}{c|}{{\color[HTML]{000000} 1.70}}                                                         & {\color[HTML]{000000} 0.32}                                                        \\
{\color[HTML]{000000} exec}                                & {\color[HTML]{000000} 253}                            & {\color[HTML]{000000} 46.25}                                                                           & {\color[HTML]{000000} 26.88}                                                       & {\color[HTML]{000000} 4.35}                                                        & {\color[HTML]{000000} 22.53}                                                       & \multicolumn{1}{c|}{{\color[HTML]{000000} 49.41}}                                                        & \multicolumn{1}{c|}{{\color[HTML]{000000} 1.58}}                                                        & \multicolumn{1}{c|}{{\color[HTML]{000000} 1.19}}                                                         & \multicolumn{1}{c|}{{\color[HTML]{000000} 0}}                                                            & \multicolumn{1}{c|}{{\color[HTML]{000000} 1.58}}                                                         & {\color[HTML]{000000} 4.29}                                                        \\
{\color[HTML]{000000} fast-pfor}                           & {\color[HTML]{000000} 754}                            & {\color[HTML]{000000} 46.02}                                                                           & {\color[HTML]{000000} 0}                                                           & {\color[HTML]{000000} 0}                                                           & {\color[HTML]{000000} 53.98}                                                       & \multicolumn{1}{c|}{{\color[HTML]{000000} 38.20}}                                                        & \multicolumn{1}{c|}{{\color[HTML]{000000} 0.93}}                                                        & \multicolumn{1}{c|}{{\color[HTML]{000000} 6.63}}                                                         & \multicolumn{1}{c|}{{\color[HTML]{000000} 0.80}}                                                         & \multicolumn{1}{c|}{{\color[HTML]{000000} 7.43}}                                                         & {\color[HTML]{000000} 4.88}                                                        \\
{\color[HTML]{000000} fileupload}                          & {\color[HTML]{000000} 168}                            & {\color[HTML]{000000} 85.12}                                                                           & {\color[HTML]{000000} 10.71}                                                       & {\color[HTML]{000000} 0.60}                                                        & {\color[HTML]{000000} 3.57}                                                        & \multicolumn{1}{c|}{{\color[HTML]{000000} 8.93}}                                                         & \multicolumn{1}{c|}{{\color[HTML]{000000} 1.19}}                                                        & \multicolumn{1}{c|}{{\color[HTML]{000000} 2.98}}                                                         & \multicolumn{1}{c|}{{\color[HTML]{000000} 1.19}}                                                         & \multicolumn{1}{c|}{{\color[HTML]{000000} 0.60}}                                                         & {\color[HTML]{000000} 23.08}                                                       \\
{\color[HTML]{000000} graph}                               & {\color[HTML]{000000} 556}                            & {\color[HTML]{000000} 40.65}                                                                           & {\color[HTML]{000000} 0}                                                           & {\color[HTML]{000000} 0}                                                           & {\color[HTML]{000000} 59.35}                                                       & \multicolumn{1}{c|}{{\color[HTML]{000000} 56.12}}                                                        & \multicolumn{1}{c|}{{\color[HTML]{000000} 0}}                                                           & \multicolumn{1}{c|}{{\color[HTML]{000000} 1.98}}                                                         & \multicolumn{1}{c|}{{\color[HTML]{000000} 0.36}}                                                         & \multicolumn{1}{c|}{{\color[HTML]{000000} 0.90}}                                                         & {\color[HTML]{000000} 5.48}                                                        \\
{\color[HTML]{000000} jansi}                               & {\color[HTML]{000000} 314}                            & {\color[HTML]{000000} 69.75}                                                                           & {\color[HTML]{000000} 0}                                                           & {\color[HTML]{000000} 0}                                                           & {\color[HTML]{000000} 30.25}                                                       & \multicolumn{1}{c|}{{\color[HTML]{000000} 29.94}}                                                        & \multicolumn{1}{c|}{{\color[HTML]{000000} 0}}                                                           & \multicolumn{1}{c|}{{\color[HTML]{000000} 0.32}}                                                         & \multicolumn{1}{c|}{{\color[HTML]{000000} 0}}                                                            & \multicolumn{1}{c|}{{\color[HTML]{000000} 0}}                                                            & {\color[HTML]{000000} 0}                                                           \\
{\color[HTML]{000000} pool}                                & {\color[HTML]{000000} 574}                            & {\color[HTML]{000000} 70.56}                                                                           & {\color[HTML]{000000} 0}                                                           & {\color[HTML]{000000} 0}                                                           & {\color[HTML]{000000} 29.44}                                                       & \multicolumn{1}{c|}{{\color[HTML]{000000} 27.00}}                                                        & \multicolumn{1}{c|}{{\color[HTML]{000000} 1.39}}                                                        & \multicolumn{1}{c|}{{\color[HTML]{000000} 0}}                                                            & \multicolumn{1}{c|}{{\color[HTML]{000000} 0.35}}                                                         & \multicolumn{1}{c|}{{\color[HTML]{000000} 0.70}}                                                         & {\color[HTML]{000000} 5.48}                                                        \\
{\color[HTML]{000000} validator}                           & {\color[HTML]{000000} 623}                            & {\color[HTML]{000000} 33.39}                                                                           & {\color[HTML]{000000} 18.78}                                                       & {\color[HTML]{000000} 20.06}                                                       & {\color[HTML]{000000} 27.77}                                                       & \multicolumn{1}{c|}{{\color[HTML]{000000} 66.29}}                                                        & \multicolumn{1}{c|}{{\color[HTML]{000000} 0}}                                                           & \multicolumn{1}{c|}{{\color[HTML]{000000} 0}}                                                            & \multicolumn{1}{c|}{{\color[HTML]{000000} 0.32}}                                                         & \multicolumn{1}{c|}{{\color[HTML]{000000} 0}}                                                            & {\color[HTML]{000000} 0.43}                                                        \\ \hline
{\color[HTML]{000000} \textbf{Total}}                      & {\color[HTML]{000000} 4431}                           & {\color[HTML]{000000} 40.01}                                                                           & {\color[HTML]{000000} 4.58}                                                        & {\color[HTML]{000000} 3.09}                                                        & {\color[HTML]{000000} 52.31}                                                       & \multicolumn{1}{c|}{{\color[HTML]{000000} 52.79}}                                                        & \multicolumn{1}{c|}{{\color[HTML]{000000} 0.81}}                                                        & \multicolumn{1}{c|}{{\color[HTML]{000000} 2.57}}                                                         & \multicolumn{1}{c|}{{\color[HTML]{000000} 0.86}}                                                         & \multicolumn{1}{c|}{{\color[HTML]{000000} 2.96}}                                                         & {\color[HTML]{000000} 3.05}                                                        \\ \hline
\end{tabular}
    \label{table:rq5-ablation}
    \vspace{-10pt} % Adjust for whitespace below
\end{table}

%\ali{added a new table for baseline results under this RQ. So in total we did 3 ablation studies. 1) program transformation. 2) program decomposition (baseline study), and 4) choice of LLM.}
%\reyhan{Ali, this section requires major rework, e.g., elaborating more on the added table and cost analysis based on the reported numbers. If we now have three ablations, the RQ description at the beginning of the section should also be updated (accordingly, we can add a few sentence at the end of introduction to discuss these new results.)}

\rebuttal{We performed three ablation studies to investigate the impact of program transformation, choice of LLM, and program decomposition on the performance of \approach.}

\subsubsection{Impact of Program Transformation} \rebuttal{We removed the program transformation component of \approach and executed the entire pipeline. The results in Table~\ref{table:rq5-ablation} show that without transformation (e.g., resolving method/constructor overloading), the performance of \approach drops  significantly: \emph{GS}, \emph{ATP}, and \emph{TPR} values decreased to $4.58\%$ (from $24.50\%$), $0.81\%$ (from $2.88\%$), and $3.05\%$ (from $9.76\%$), respectively. This is because Python does not support overloading and only considers the last method/constructor implementation, resulting in runtime errors or test failures. Also, \emph{GE} values increase due to the interference of overloaded code constructs with GraalVM.}

\begin{table}[t]
    \setlength{\tabcolsep}{1.3pt}
    \scriptsize
    \centering
    \vspace{-5pt}
    \caption{\rebuttal{Effectiveness of \approach with GPT-4o in compositional translation and validation. Abbreviations in the table stand for \textbf{AMF}: \#Application Method Fragments, \textbf{SNEF}: Source Non-Exercised Fragments, \textbf{GS}: Graal Success, \textbf{GF}: Graal Fail, \textbf{GE}: Graal Error, \textbf{TNEF}: Target Non-Exercised Fragments, \textbf{ATP}: Fragments All Test Pass, \textbf{OTF}: Fragments One Test Fail, \textbf{MTF}: Fragments Many Test Fail, \textbf{ATF}: Fragments All Test Fail, \textbf{TPR}: Test Pass Rate, \textbf{O}: Overall, \textbf{RE}: Runtime Error, \textbf{AF}: Assertion Failure, and \textbf{M1}: Number of \textit{AMFs} that GraalVM could not execute (\textit{GE}) but translated test fragments exercised.}}
    % \reyhan{Ali, please change the color of text in this table to blue. Also, remove column AMF (as the numbers are not changed) and add a column at the end showing the cost of using gpt-4o}
    \vspace{-5pt}
    \begin{tabular}{l|c|c|ccc|cccccccccccc|cc|c}
\hline
{\color[HTML]{000000} }                                    & {\color[HTML]{000000} }                                                                                         & {\color[HTML]{000000} }                                                                                & \multicolumn{3}{c|}{{\color[HTML]{000000} \textbf{GraalVM}}}                                                                                                                      & \multicolumn{12}{c|}{{\color[HTML]{000000} \textbf{Test Translation}}}                                                                                                                                                                                                                                                                                                                                                                                                                                                                                                                                                                                                                                                                                                                                                                                                                               & \multicolumn{2}{c|}{{\color[HTML]{000000} }}                                                    & {\color[HTML]{000000} }                                                                               \\ \cline{4-18}
{\color[HTML]{000000} }                                    & {\color[HTML]{000000} }                                                                                         & {\color[HTML]{000000} }                                                                                & {\color[HTML]{000000} }                                   & {\color[HTML]{000000} }                                   & {\color[HTML]{000000} }                                   & \multicolumn{1}{c|}{{\color[HTML]{000000} }}                                                                               & \multicolumn{1}{c|}{{\color[HTML]{000000} }}                                                                              & \multicolumn{3}{c|}{{\color[HTML]{000000} \textbf{OTF (\%)}}}                                                                                                              & \multicolumn{3}{c|}{{\color[HTML]{000000} \textbf{MTF (\%)}}}                                                                                                              & \multicolumn{3}{c|}{{\color[HTML]{000000} \textbf{ATF (\%)}}}                                                                                                              & {\color[HTML]{000000} }                                                                              & \multicolumn{2}{c|}{\multirow{-2}{*}{{\color[HTML]{000000} \textbf{M1}}}}                       & {\color[HTML]{000000} }                                                                               \\ \cline{9-17} \cline{19-20}
\multirow{-3}{*}{{\color[HTML]{000000} \textbf{Subjects}}} & \multirow{-3}{*}{{\color[HTML]{000000} \textbf{\begin{tabular}[c]{@{}c@{}}Syntax\\ Check\\ (\%)\end{tabular}}}} & \multirow{-3}{*}{{\color[HTML]{000000} \textbf{\begin{tabular}[c]{@{}c@{}}SNEF\\  (\%)\end{tabular}}}} & \multirow{-2}{*}{{\color[HTML]{000000} \textbf{GS (\%)}}} & \multirow{-2}{*}{{\color[HTML]{000000} \textbf{GF (\%)}}} & \multirow{-2}{*}{{\color[HTML]{000000} \textbf{GE (\%)}}} & \multicolumn{1}{c|}{\multirow{-2}{*}{{\color[HTML]{000000} \textbf{\begin{tabular}[c]{@{}c@{}}TNEF\\ (\%)\end{tabular}}}}} & \multicolumn{1}{c|}{\multirow{-2}{*}{{\color[HTML]{000000} \textbf{\begin{tabular}[c]{@{}c@{}}ATP\\ (\%)\end{tabular}}}}} & \multicolumn{1}{c|}{{\color[HTML]{000000} \textbf{O}}} & \multicolumn{1}{c|}{{\color[HTML]{000000} \textbf{RE}}} & \multicolumn{1}{c|}{{\color[HTML]{000000} \textbf{AF}}} & \multicolumn{1}{c|}{{\color[HTML]{000000} \textbf{O}}} & \multicolumn{1}{c|}{{\color[HTML]{000000} \textbf{RE}}} & \multicolumn{1}{c|}{{\color[HTML]{000000} \textbf{AF}}} & \multicolumn{1}{c|}{{\color[HTML]{000000} \textbf{O}}} & \multicolumn{1}{c|}{{\color[HTML]{000000} \textbf{RE}}} & \multicolumn{1}{c|}{{\color[HTML]{000000} \textbf{AF}}} & \multirow{-2}{*}{{\color[HTML]{000000} \textbf{\begin{tabular}[c]{@{}c@{}}TPR\\ (\%)\end{tabular}}}} & \multicolumn{1}{c|}{{\color[HTML]{000000} \textbf{All}}} & {\color[HTML]{000000} \textbf{Some}} & \multirow{-3}{*}{{\color[HTML]{000000} \textbf{\begin{tabular}[c]{@{}c@{}}Cost\\ (\$)\end{tabular}}}} \\ \hline
{\color[HTML]{000000} cli}                                 & {\color[HTML]{000000} 99.63}                                                                                    & {\color[HTML]{000000} 5.86}                                                                            & {\color[HTML]{000000} 76.92}                              & {\color[HTML]{000000} 8.79}                               & {\color[HTML]{000000} 8.42}                               & \multicolumn{1}{c|}{{\color[HTML]{000000} 78.39}}                                                                          & \multicolumn{1}{c|}{{\color[HTML]{000000} 4.76}}                                                                          & {\color[HTML]{000000} 3.66}                            & {\color[HTML]{000000} 90.00}                            & \multicolumn{1}{c|}{{\color[HTML]{000000} 10.00}}       & {\color[HTML]{000000} 6.23}                            & {\color[HTML]{000000} 100}                              & \multicolumn{1}{c|}{{\color[HTML]{000000} 0}}           & {\color[HTML]{000000} 1.10}                            & {\color[HTML]{000000} 100}                              & \multicolumn{1}{c|}{{\color[HTML]{000000} 0}}           & {\color[HTML]{000000} 2.47}                                                                          & {\color[HTML]{000000} 0}                                 & {\color[HTML]{000000} 1}             & {\color[HTML]{000000} 19.69}                                                                          \\
{\color[HTML]{000000} codec}                               & {\color[HTML]{000000} 97.79}                                                                                    & {\color[HTML]{000000} 8.97}                                                                            & {\color[HTML]{000000} 42.50}                              & {\color[HTML]{000000} 28.53}                              & {\color[HTML]{000000} 20.00}                              & \multicolumn{1}{c|}{{\color[HTML]{000000} 74.56}}                                                                          & \multicolumn{1}{c|}{{\color[HTML]{000000} 3.82}}                                                                          & {\color[HTML]{000000} 4.12}                            & {\color[HTML]{000000} 78.57}                            & \multicolumn{1}{c|}{{\color[HTML]{000000} 21.43}}       & {\color[HTML]{000000} 6.18}                            & {\color[HTML]{000000} 30.62}                            & \multicolumn{1}{c|}{{\color[HTML]{000000} 69.38}}       & {\color[HTML]{000000} 2.35}                            & {\color[HTML]{000000} 73.33}                            & \multicolumn{1}{c|}{{\color[HTML]{000000} 26.67}}       & {\color[HTML]{000000} 9.57}                                                                          & {\color[HTML]{000000} 2}                                 & {\color[HTML]{000000} 18}            & {\color[HTML]{000000} 39.97}                                                                          \\
{\color[HTML]{000000} csv}                                 & {\color[HTML]{000000} 98.30}                                                                                    & {\color[HTML]{000000} 9.36}                                                                            & {\color[HTML]{000000} 41.28}                              & {\color[HTML]{000000} 25.96}                              & {\color[HTML]{000000} 23.40}                              & \multicolumn{1}{c|}{{\color[HTML]{000000} 86.81}}                                                                          & \multicolumn{1}{c|}{{\color[HTML]{000000} 0}}                                                                             & {\color[HTML]{000000} 0}                               & {\color[HTML]{000000} 0}                                & \multicolumn{1}{c|}{{\color[HTML]{000000} 0}}           & {\color[HTML]{000000} 1.70}                            & {\color[HTML]{000000} 78.17}                            & \multicolumn{1}{c|}{{\color[HTML]{000000} 21.83}}       & {\color[HTML]{000000} 2.13}                            & {\color[HTML]{000000} 96.72}                            & \multicolumn{1}{c|}{{\color[HTML]{000000} 3.28}}        & {\color[HTML]{000000} 0.98}                                                                          & {\color[HTML]{000000} 0}                                 & {\color[HTML]{000000} 8}             & {\color[HTML]{000000} 16.66}                                                                          \\
{\color[HTML]{000000} exec}                                & {\color[HTML]{000000} 99.60}                                                                                    & {\color[HTML]{000000} 45.16}                                                                           & {\color[HTML]{000000} 35.89}                              & {\color[HTML]{000000} 2.42}                               & {\color[HTML]{000000} 16.53}                              & \multicolumn{1}{c|}{{\color[HTML]{000000} 44.76}}                                                                          & \multicolumn{1}{c|}{{\color[HTML]{000000} 4.84}}                                                                          & {\color[HTML]{000000} 0}                               & {\color[HTML]{000000} 0}                                & \multicolumn{1}{c|}{{\color[HTML]{000000} 0}}           & {\color[HTML]{000000} 5.24}                            & {\color[HTML]{000000} 0}                                & \multicolumn{1}{c|}{{\color[HTML]{000000} 100}}         & {\color[HTML]{000000} 0}                               & {\color[HTML]{000000} 0}                                & \multicolumn{1}{c|}{{\color[HTML]{000000} 0}}           & {\color[HTML]{000000} 28.62}                                                                         & {\color[HTML]{000000} 1}                                 & {\color[HTML]{000000} 1}             & {\color[HTML]{000000} 3.90}                                                                           \\
{\color[HTML]{000000} fast-pfor}                           & {\color[HTML]{000000} 97.46}                                                                                    & {\color[HTML]{000000} 45.45}                                                                           & {\color[HTML]{000000} 14.71}                              & {\color[HTML]{000000} 16.98}                              & {\color[HTML]{000000} 22.86}                              & \multicolumn{1}{c|}{{\color[HTML]{000000} 52.94}}                                                                          & \multicolumn{1}{c|}{{\color[HTML]{000000} 1.47}}                                                                          & {\color[HTML]{000000} 0}                               & {\color[HTML]{000000} 0}                                & \multicolumn{1}{c|}{{\color[HTML]{000000} 0}}           & {\color[HTML]{000000} 0}                               & {\color[HTML]{000000} 0}                                & \multicolumn{1}{c|}{{\color[HTML]{000000} 0}}           & {\color[HTML]{000000} 0.13}                            & {\color[HTML]{000000} 100}                              & \multicolumn{1}{c|}{{\color[HTML]{000000} 0}}           & {\color[HTML]{000000} 2.41}                                                                          & {\color[HTML]{000000} 2}                                 & {\color[HTML]{000000} 0}             & {\color[HTML]{000000} 12.93}                                                                          \\
{\color[HTML]{000000} fileupload}                          & {\color[HTML]{000000} 99.48}                                                                                    & {\color[HTML]{000000} 86.98}                                                                           & {\color[HTML]{000000} 9.90}                               & {\color[HTML]{000000} 0.52}                               & {\color[HTML]{000000} 2.60}                               & \multicolumn{1}{c|}{{\color[HTML]{000000} 4.17}}                                                                           & \multicolumn{1}{c|}{{\color[HTML]{000000} 6.77}}                                                                          & {\color[HTML]{000000} 1.56}                            & {\color[HTML]{000000} 0}                                & \multicolumn{1}{c|}{{\color[HTML]{000000} 100}}         & {\color[HTML]{000000} 0.52}                            & {\color[HTML]{000000} 100}                              & \multicolumn{1}{c|}{{\color[HTML]{000000} 0}}           & {\color[HTML]{000000} 0}                               & {\color[HTML]{000000} 0}                                & \multicolumn{1}{c|}{{\color[HTML]{000000} 0}}           & {\color[HTML]{000000} 54.84}                                                                         & {\color[HTML]{000000} 1}                                 & {\color[HTML]{000000} 3}             & {\color[HTML]{000000} 2.25}                                                                           \\
{\color[HTML]{000000} graph}                               & {\color[HTML]{000000} 98.71}                                                                                    & {\color[HTML]{000000} 41.22}                                                                           & {\color[HTML]{000000} 27.73}                              & {\color[HTML]{000000} 19.04}                              & {\color[HTML]{000000} 12.01}                              & \multicolumn{1}{c|}{{\color[HTML]{000000} 58.78}}                                                                          & \multicolumn{1}{c|}{{\color[HTML]{000000} 0}}                                                                             & {\color[HTML]{000000} 0}                               & {\color[HTML]{000000} 0}                                & \multicolumn{1}{c|}{{\color[HTML]{000000} 0}}           & {\color[HTML]{000000} 0}                               & {\color[HTML]{000000} 0}                                & \multicolumn{1}{c|}{{\color[HTML]{000000} 0}}           & {\color[HTML]{000000} 0}                               & {\color[HTML]{000000} 0}                                & \multicolumn{1}{c|}{{\color[HTML]{000000} 0}}           & {\color[HTML]{000000} 0}                                                                             & {\color[HTML]{000000} 0}                                 & {\color[HTML]{000000} 0}             & {\color[HTML]{000000} 12.03}                                                                          \\
{\color[HTML]{000000} jansi}                               & {\color[HTML]{000000} 99.76}                                                                                    & {\color[HTML]{000000} 76.53}                                                                           & {\color[HTML]{000000} 8.07}                               & {\color[HTML]{000000} 11.98}                              & {\color[HTML]{000000} 3.42}                               & \multicolumn{1}{c|}{{\color[HTML]{000000} 23.47}}                                                                          & \multicolumn{1}{c|}{{\color[HTML]{000000} 0}}                                                                             & {\color[HTML]{000000} 0}                               & {\color[HTML]{000000} 0}                                & \multicolumn{1}{c|}{{\color[HTML]{000000} 0}}           & {\color[HTML]{000000} 0}                               & {\color[HTML]{000000} 0}                                & \multicolumn{1}{c|}{{\color[HTML]{000000} 0}}           & {\color[HTML]{000000} 0}                               & {\color[HTML]{000000} 0}                                & \multicolumn{1}{c|}{{\color[HTML]{000000} 0}}           & {\color[HTML]{000000} 0}                                                                             & {\color[HTML]{000000} 0}                                 & {\color[HTML]{000000} 0}             & {\color[HTML]{000000} 3.66}                                                                           \\
{\color[HTML]{000000} pool}                                & {\color[HTML]{000000} 99.71}                                                                                    & {\color[HTML]{000000} 77.71}                                                                           & {\color[HTML]{000000} 7.48}                               & {\color[HTML]{000000} 1.32}                               & {\color[HTML]{000000} 13.49}                              & \multicolumn{1}{c|}{{\color[HTML]{000000} 21.70}}                                                                          & \multicolumn{1}{c|}{{\color[HTML]{000000} 0.59}}                                                                          & {\color[HTML]{000000} 0}                               & {\color[HTML]{000000} 0}                                & \multicolumn{1}{c|}{{\color[HTML]{000000} 0}}           & {\color[HTML]{000000} 0}                               & {\color[HTML]{000000} 0}                                & \multicolumn{1}{c|}{{\color[HTML]{000000} 0}}           & {\color[HTML]{000000} 0}                               & {\color[HTML]{000000} 0}                                & \multicolumn{1}{c|}{{\color[HTML]{000000} 0}}           & {\color[HTML]{000000} 2.09}                                                                          & {\color[HTML]{000000} 0}                                 & {\color[HTML]{000000} 0}             & {\color[HTML]{000000} 7.33}                                                                           \\
{\color[HTML]{000000} validator}                           & {\color[HTML]{000000} 99.23}                                                                                    & {\color[HTML]{000000} 36.69}                                                                           & {\color[HTML]{000000} 38.24}                              & {\color[HTML]{000000} 15.94}                              & {\color[HTML]{000000} 9.13}                               & \multicolumn{1}{c|}{{\color[HTML]{000000} 54.95}}                                                                          & \multicolumn{1}{c|}{{\color[HTML]{000000} 4.02}}                                                                          & {\color[HTML]{000000} 2.32}                            & {\color[HTML]{000000} 93.33}                            & \multicolumn{1}{c|}{{\color[HTML]{000000} 6.67}}        & {\color[HTML]{000000} 1.55}                            & {\color[HTML]{000000} 70.59}                            & \multicolumn{1}{c|}{{\color[HTML]{000000} 29.41}}       & {\color[HTML]{000000} 0.46}                            & {\color[HTML]{000000} 100}                              & \multicolumn{1}{c|}{{\color[HTML]{000000} 0}}           & {\color[HTML]{000000} 10.41}                                                                         & {\color[HTML]{000000} 0}                                 & {\color[HTML]{000000} 3}             & {\color[HTML]{000000} 25.53}                                                                          \\ \hline
{\color[HTML]{000000} \textbf{Total}}                      & {\color[HTML]{000000} 98.80}                                                                                    & {\color[HTML]{000000} 43.43}                                                                           & {\color[HTML]{000000} 27.83}                              & {\color[HTML]{000000} 14.54}                              & {\color[HTML]{000000} 14.20}                              & \multicolumn{1}{c|}{{\color[HTML]{000000} 50.64}}                                                                          & \multicolumn{1}{c|}{{\color[HTML]{000000} 2.26}}                                                                          & {\color[HTML]{000000} 1.20}                            & {\color[HTML]{000000} 80.36}                            & \multicolumn{1}{c|}{{\color[HTML]{000000} 19.64}}       & {\color[HTML]{000000} 1.87}                            & {\color[HTML]{000000} 59.39}                            & \multicolumn{1}{c|}{{\color[HTML]{000000} 40.61}}       & {\color[HTML]{000000} 0.60}                            & {\color[HTML]{000000} 86.59}                            & \multicolumn{1}{c|}{{\color[HTML]{000000} 13.41}}       & {\color[HTML]{000000} 6.45}                                                                          & {\color[HTML]{000000} 6}                                 & {\color[HTML]{000000} 34}            & {\color[HTML]{000000} 143.95}                                                                         \\ \hline
\end{tabular}
    \label{table:rq5-merged-effectiveness}
    \vspace{-13pt}
\end{table}

% Total AMF = 4654
% GS = 1295
% M1 All = 6
% M1 Some = 34
% runtime validation = (GS + M1 Some) / Total AMF = (1295 + 34) / 4654 = 28.56 %

% functional equivalence = (GS + M1 All) / Total AMF = (1295 + 6) / 4654 = 27.95 %
% \ali{adjust formulas: Runtime=GS+M1Some. functional=GS+M1 All}

\subsubsection{Choice of LLM} 
\begin{wrapfigure}{R}{0.27\linewidth}
  \centering
  \scriptsize
  \begin{minipage}{\linewidth}
  \vspace{-30pt} % to remove the whitespace above
    \includegraphics[width=\linewidth]{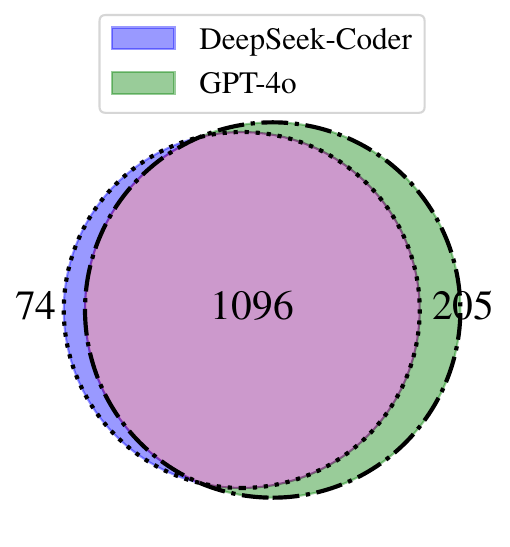}
    \vspace{-25pt}
    \caption{Functional equivalence overlap of AMFs (Graal Success) between LLMs.}
    \label{fig:venn-diagram}
  \vspace{-22pt} % to remove the whitespace below
\end{minipage}
\end{wrapfigure}

\rebuttal{For this experiment, we replaced the \llm with \gptfo and repeated the entire pipeline of \approach (Table~\ref{table:rq5-merged-effectiveness}). A stronger model such as \gptfo improves the translation quality---functional equivalence increases from $25.14\%$ to $27.95\%$. For some projects, the ATP and TPR rates are higher for \llm translations, whereas for the others, \gptfo results in higher values. We investigated each LLM's successful AMF translations to better understand the differences. We observed a huge overlap between successful AMFs and the unique benefits each LLM provides in code translation (Figure~\ref{fig:venn-diagram}). \gptfo handles API translation and type casting better, resolving the first three translation bugs discussed in \S\ref{sec:rq-bugs}. In contrast, it tends to add unnecessary code, mostly due to error handling, which results in a functional mismatch. In the example below, \texttt{\small{create2}} method can take a \texttt{\small None} value, and its implementation performs error handling when necessary. \gptfo adds unnecessary error handling code, interfering with program logic and resulting in test failures.}
% \reyhan{placeholder for numbers should be updated similar to table 2}
%Overall, \gptfo handles API translation and type casting better, resolving the first three translation bugs discussed in \S\ref{sec:rq-bugs}. Conversely, it can incorrectly translate some simple cases during code and test translation due to extra error handling, as shown in the code snippet below.} \reyhan{placeholder for numbers should be updated similar to table 2}
% \saurabh{We might want to rephrase ``over-creative''; also providing an example would help}. \ali{Not sure if we have space to add an example, but we can change over-creative to ``injecting extra error-handling''.}

\vspace{5pt}
\noindent
\begin{minipage}{.475\linewidth}
% \begin{lstlisting}[language = source, columns=fullflexible, basicstyle=\scriptsize\ttfamily,showlines=true]
\begin{minted}[frame=lines,framesep=1mm,baselinestretch=0.5, fontsize=\scriptsize, breaklines, breakanywhere, linenos,numbersep=2pt]{java}
        ----------- JAVA SOURCE CODE -----------
public Option create1(OptionGroup optGroup) {
    return create2(optGroup);

}
\end{minted}
% \end{lstlisting}
\end{minipage}\hfill
\begin{minipage}{.475\linewidth}
% \begin{lstlisting}[language = source, columns=fullflexible, basicstyle=\scriptsize\ttfamily,showlines=true]
\begin{minted}[escapeinside=||, frame=lines,framesep=1mm,baselinestretch=0.5, fontsize=\scriptsize, breaklines, breakanywhere, linenos,numbersep=2pt, highlightlines={3,4}]{py}
        ---------- PYTHON TRANSLATION ----------
def create1(self, optGroup: OptionGroup) -> Option:
|\xglobal\colorlet{FancyVerbHighlightColor}{cambridgeblue}|+   if optGroup is None:
|\xglobal\colorlet{FancyVerbHighlightColor}{cambridgeblue}|+       raise ValueError("optGroup is None")
    return self.create2(optGroup)
\end{minted}
% \end{lstlisting}
\end{minipage}
\vspace{5pt}

\rebuttal{It is worth noting that using commercial LLMs comes at a cost. The last column of Table~\ref{table:rq5-merged-effectiveness} (column \emph{Cost}) shows the cost of using \gptfo for repeating the experiments, resulting in the total cost of $\$143.95$ for translating all the subjects (average cost of $\$14.39$ per project).}

\subsubsection{Impact of Program Decomposition} \rebuttal{For this ablation study, we prompted \gptfo and \llm \emph{file-by-file} and evaluated translation correctness through test execution and GraalVM (Table~\ref{table:rq5-baseline}). Not surprisingly, a considerable percentage of the files exceeded the model context window size, particularly for DeepSeekCoder, with $9.08\%$ of the files encountering this problem. Among the prompted files, $21.36\%$ ($19.44\%$ for DeepSeekCoder and $1.92\%$ for \gptfo) were syntactically incorrect. Translations that passed the syntactic correctness check did not pass any translated test execution, whereas GraalVM validated $2.56\%$ of the files in total ($1.02\%$ for DeepSeekCoder and $1.54\%$ for \gptfo). Note that file-level GraalVM validation does not mean that all the methods are correctly translated and validated---only the methods in the class that are executed by tests. Manual analysis of the results of this experiment revealed that one prominent culprit of test failures was LLM hallucination with method/variable names. Given that we had no skeleton construction in this baseline, such issues could not be avoided; this demonstrates the usefulness of skeleton construction and incremental translation in \approach.}

\mybox{\textnormal{\textbf{Summary.} \rebuttal{Omitting program transformation and program decomposition significantly lowers the effectiveness of \approach.
%, with \textit{GS} decreasing from $24.50\%$ to $4.58\%$, whereas u
A stronger model such as \gptfo resolves non-trivial issues concerning type casting and API translation but may result in trivial translation bugs. When possible, users of \approach can prompt multiple LLMs to achieve better translation performance. 
%increases the functional equivalence from $...\%$ to $...\%$. %Moreover, file-by-file translation of projects achieves very low success rate, with no passing translated tests and only $2.4\%$ Graal success.
}}}

\section{Related Work}
\label{sec:related-work}

There are generally two main categories of techniques for translating code from one PL to another: (1) using transpilers and statistical machine translation and (2) leveraging language models.

\subsection{Code Translation Using Non-LLM-Based Approaches} 
%In this domain, t
Tools like C2Rust~\cite{c2rust}, CxGo~\cite{c2go}, Sharpen~\cite{sharpen}, and Java2CSharp~\cite{java2csharp} 
%have been developed to translate 
translate code from C to Rust, C to Go, and Java to C\# respectively. 
%However, a recent study~\cite{pan2024lost} revealed that, apart from C2Rust and CxGo, other tools lack proper maintenance. For CxGo, language models outperform traditional approaches, whereas for C2Rust, language models generate safer but less effective code, aligning with the primary goal of translating C code to Rust. 
%In terms of s
A series of statistical machine translation techniques 
%, works by Nguyen ~\textit{et al.}~
\cite{nguyen2013lexical,nguyen2014migrating,nguyen2015divide,chen2018tree} 
%and Chen~\textit{et al.}~\cite{chen2018tree} 
focus on translating Java to C\#. Deep learning approaches have also been applied for code translation~\cite{roziere2020unsupervised, roziere2021leveraging}. None of these efforts have tackled translating real-world Java projects to Python. LLM-based techniques are also superior to transpilers in terms of performance or readability~\cite{pan2024lost}. 

\begin{table}[t]
    \scriptsize
    \centering
    \vspace{-5pt}
    \caption{\rebuttal{Importance of program decomposition. Abbreviations stand for \textbf{GS}: Graal Success and \textbf{TPR}: Test Pass Rate. Since Java can contain multiple classes in one file, \textbf{\#Files} is smaller from \textbf{\#Classes} in Table~\ref{table:stats}.}}
    % \reyhan{change the color of the text in table blue and remove the cost column. I'm confused where the cost numbers are coming from: are these estimates or are these from the experiments you did running the gpt4-o for the ablation? for table 5, we should use the numbers that you ran experiments on not estimation}}
    \vspace{-5pt}
    \begin{tabular}{l|c|cccc|cccc}
\hline
{\color[HTML]{000000} }                                    & {\color[HTML]{000000} }                                    & \multicolumn{4}{c|}{{\color[HTML]{000000} \textbf{GPT-4o}}}                                                                                                                                                                                                                                               & \multicolumn{4}{c}{{\color[HTML]{000000} \textbf{DeepSeek-Coder}}}                                                                                                                                                                                                                                        \\ \cline{3-10} 
\multirow{-2}{*}{{\color[HTML]{000000} \textbf{Subjects}}} & \multirow{-2}{*}{{\color[HTML]{000000} \textbf{\# Files}}} & {\color[HTML]{000000} \textbf{\begin{tabular}[c]{@{}c@{}}Over\\ Context\end{tabular}}} & {\color[HTML]{000000} \textbf{\begin{tabular}[c]{@{}c@{}}Syntax\\ Error\end{tabular}}} & {\color[HTML]{000000} \textbf{GS}} & {\color[HTML]{000000} \textbf{\begin{tabular}[c]{@{}c@{}}TPR\\ (\%)\end{tabular}}} & {\color[HTML]{000000} \textbf{\begin{tabular}[c]{@{}c@{}}Over\\ Context\end{tabular}}} & {\color[HTML]{000000} \textbf{\begin{tabular}[c]{@{}c@{}}Syntax\\ Error\end{tabular}}} & {\color[HTML]{000000} \textbf{GS}} & {\color[HTML]{000000} \textbf{\begin{tabular}[c]{@{}c@{}}TPR\\ (\%)\end{tabular}}} \\ \hline
{\color[HTML]{000000} cli}                                 & {\color[HTML]{000000} 51}                                  & {\color[HTML]{000000} 0}                                                               & {\color[HTML]{000000} 0}                                                               & {\color[HTML]{000000} 1}           & {\color[HTML]{000000} 0}                                                           & {\color[HTML]{000000} 5}                                                               & {\color[HTML]{000000} 6}                                                               & {\color[HTML]{000000} 0}           & {\color[HTML]{000000} 0}                                                           \\
{\color[HTML]{000000} codec}                               & {\color[HTML]{000000} 136}                                 & {\color[HTML]{000000} 0}                                                               & {\color[HTML]{000000} 1}                                                               & {\color[HTML]{000000} 7}           & {\color[HTML]{000000} 0}                                                           & {\color[HTML]{000000} 28}                                                              & {\color[HTML]{000000} 46}                                                              & {\color[HTML]{000000} 4}           & {\color[HTML]{000000} 0}                                                           \\
{\color[HTML]{000000} csv}                                 & {\color[HTML]{000000} 33}                                  & {\color[HTML]{000000} 0}                                                               & {\color[HTML]{000000} 3}                                                               & {\color[HTML]{000000} 0}           & {\color[HTML]{000000} 0}                                                           & {\color[HTML]{000000} 6}                                                               & {\color[HTML]{000000} 8}                                                               & {\color[HTML]{000000} 0}           & {\color[HTML]{000000} 0}                                                           \\
{\color[HTML]{000000} exec}                                & {\color[HTML]{000000} 54}                                  & {\color[HTML]{000000} 0}                                                               & {\color[HTML]{000000} 0}                                                               & {\color[HTML]{000000} 0}           & {\color[HTML]{000000} 0}                                                           & {\color[HTML]{000000} 1}                                                               & {\color[HTML]{000000} 4}                                                               & {\color[HTML]{000000} 0}           & {\color[HTML]{000000} 0}                                                           \\
{\color[HTML]{000000} fast-pfor}                           & {\color[HTML]{000000} 85}                                  & {\color[HTML]{000000} 1}                                                               & {\color[HTML]{000000} 1}                                                               & {\color[HTML]{000000} 1}           & {\color[HTML]{000000} 0}                                                           & {\color[HTML]{000000} 12}                                                              & {\color[HTML]{000000} 30}                                                              & {\color[HTML]{000000} 1}           & {\color[HTML]{000000} 0}                                                           \\
{\color[HTML]{000000} fileupload}                          & {\color[HTML]{000000} 43}                                  & {\color[HTML]{000000} 0}                                                               & {\color[HTML]{000000} 0}                                                               & {\color[HTML]{000000} 0}           & {\color[HTML]{000000} 0}                                                           & {\color[HTML]{000000} 2}                                                               & {\color[HTML]{000000} 4}                                                               & {\color[HTML]{000000} 0}           & {\color[HTML]{000000} 0}                                                           \\
{\color[HTML]{000000} graph}                               & {\color[HTML]{000000} 159}                                 & {\color[HTML]{000000} 0}                                                               & {\color[HTML]{000000} 6}                                                               & {\color[HTML]{000000} 3}           & {\color[HTML]{000000} 0}                                                           & {\color[HTML]{000000} 0}                                                               & {\color[HTML]{000000} 17}                                                              & {\color[HTML]{000000} 3}           & {\color[HTML]{000000} 0}                                                           \\
{\color[HTML]{000000} jansi}                               & {\color[HTML]{000000} 30}                                  & {\color[HTML]{000000} 0}                                                               & {\color[HTML]{000000} 0}                                                               & {\color[HTML]{000000} 0}           & {\color[HTML]{000000} 0}                                                           & {\color[HTML]{000000} 3}                                                               & {\color[HTML]{000000} 10}                                                              & {\color[HTML]{000000} 0}           & {\color[HTML]{000000} 0}                                                           \\
{\color[HTML]{000000} pool}                                & {\color[HTML]{000000} 72}                                  & {\color[HTML]{000000} 0}                                                               & {\color[HTML]{000000} 1}                                                               & {\color[HTML]{000000} 0}           & {\color[HTML]{000000} 0}                                                           & {\color[HTML]{000000} 4}                                                               & {\color[HTML]{000000} 8}                                                               & {\color[HTML]{000000} 0}           & {\color[HTML]{000000} 0}                                                           \\
{\color[HTML]{000000} validator}                           & {\color[HTML]{000000} 119}                                 & {\color[HTML]{000000} 0}                                                               & {\color[HTML]{000000} 3}                                                               & {\color[HTML]{000000} 0}           & {\color[HTML]{000000} 0}                                                           & {\color[HTML]{000000} 10}                                                              & {\color[HTML]{000000} 19}                                                              & {\color[HTML]{000000} 0}           & {\color[HTML]{000000} 0}                                                           \\ \hline
{\color[HTML]{000000} \textbf{Total}}                      & {\color[HTML]{000000} 782}                                 & {\color[HTML]{000000} 1}                                                               & {\color[HTML]{000000} 15}                                                              & {\color[HTML]{000000} 12}          & {\color[HTML]{000000} 0}                                                           & {\color[HTML]{000000} 71}                                                              & {\color[HTML]{000000} 152}                                                             & {\color[HTML]{000000} 8}           & {\color[HTML]{000000} 0}                                                           \\ \hline
\end{tabular}
    \label{table:rq5-baseline}
    \vspace{-10pt}
\end{table}

\subsection{Code Translation Using LLMs} Recently, LLMs have been employed for code translation~\cite{pan2024lost, tipirneni2024structcoder, di2024codefuse, yan2023codetransocean, yin2024rectifier, jiao2023evaluation}, demonstrating high success rates on crafted examples but poor performance on real-world projects. Other studies~\cite{zhu2022multilingual, ahmad2021avatar} have also applied language models for code translation, mainly focusing on crafted benchmarks.
% We also noticed other parallel work
Concurrently to our work, two other techniques for repository-level code translation focusing on different language pairs were proposed~\cite{shetty2024syzygy,zhang2024scalable}. \textsc{Syzygy}~\cite{shetty2024syzygy} translates repository-level C to Rust using GPT-4. Oxidizer~\cite{zhang2024scalable} leverages language feature mapping and type-driven techniques for translating Go to Rust. Both techniques use I/O equivalence for validating their translations. There are also approaches that use transpiler output to guide LLM-based code translation~\cite{yang2024vert}. However, the limitation of such work is the availability of robust and well-maintained transpilers, which, in many cases, may not be feasible. Nitin et al.~\cite{nitin2024spectra} presented a specification-based translation, where a natural language specification is captured from the source code, which helps the translation process. Yang~et~al.~\cite{yang2024exploring} used tests to assist the translation. Compared to previous work,  our contributions include a compositional and validation-guided code translation approach that leverages two types of validation techniques and evaluation of the approach on 10 real-world projects.

% the major differences are (a) the first attempt to translate a real-world project, (b) modular translation, and (c) a validation-guided translation approach.

%\vspace{-8pt}
\section{Threats to Validity}
\label{sec:threats}
%Like most approaches, \approach possesses some limitations and comes with a list of threats to the validity. In this section, we will discuss how we mitigated various threats.

\vspace{3pt}
\textbf{External Validity.}
One of the key external threats is the generalizability of \approach. 
%To expedite the research in repository-level translation, we 
We built and evaluated the first version of \approach to translate from Java to Python. \rebuttal{However, our pipeline is generic, and with minimal effort, the current implementation can translate Java programs to more target languages (e.g., languages supported by GraalVM)}. Furthermore, the majority of the tools that we used support a large set of programming languages such as JavaScript, Ruby, C/C++, and Rust. 
%\rebuttal{Another threat is that \approach removes certain third-party libraries before translation. Translating repository-level code along with their libraries, especially when the target language does not have equivalent libraries, is a challenging problem that we plan to address in our future work.}

\vspace{3pt}
\textbf{Internal Validity.}
%We ran \approach with greedy decoding and temperature $0$, which reduces the model's creativity but makes the output consistent with several runs. 
One threat can be the manual validation of the translated types. To address that, several authors  verified the types individually and consulted API documents when necessary. \rebuttal{Another threat is that, while all successes reported by the GraalVM validation are true successes, we may have \emph{underestimated} the capabilities of \approach by a considerable margin due to the significant proportion of errors caused by limitations in the GraalVM validation approach. To mitigate this threat, we manually augmented the universal type map to support a more comprehensive translation of types. That said, we have implemented \emph{driver} code templates to provide a mechanism for adding the support for more types by the users of \approach if needed.}
%\rangeet{How to mitigate that threat?} % Mrigank (response): by supporting the translation of more types (at least those present in the project to be translated; these are small "driver" codes and can be quickly written for a given project once the type-resolution is determined).  [for impure hashing and equality functions, the graal validation script itself needs some hacks...] [we can also preserve identity of wrappers of primitive types by creating similar wrappers in Python] [can develop some heuristics to disambiguate types when there is limited type information]

\vspace{3pt}
\textbf{Construct Validity.} In order to mitigate construct validity, \approach is built and validated with well-vetted tools, such as GraalVM~\cite{graalvm}, JaCoCo~\cite{jacoco}, Python coverage~\cite{coverage-lib}, CodeQL~\cite{codeql}, etc.

%\vspace{-8pt}
\section{Concluding Remarks}
\label{sec:concluding-remarks}

In this paper, we introduced \approach, a neuro-symbolic approach that combines the power of static analysis and abilities of LLMs in code synthesis to automate repository-level code translation and validation. \approach decomposes the program into smaller fragments and translates the fragments in  reverse call order, incrementally building the source project in the target language. In addition to syntactic checks, \approach implements two types of validation through GraalVM and test translation. %Our results demonstrate the effectiveness of \approach in translating ten real-world Java projects to Python, achieving \rebuttal{$99.4\%$} syntactical correctness, \rebuttal{$48.4\%$} runtime behavior validation, and \rebuttal{$29.3\%$} functional equivalence. 
\approach is the first approach to translate and validate real-world projects, and we envision several research directions to advance repository-level code translation and validation. 

One of the major challenges in repository-level code translation is identifying suitable library APIs in the target PL. Often, equivalent Python APIs may not exist, requiring new code generation or translation of the library API itself. Even if similar libraries exist, the logic of libraries might be different in two PLs. \approach supports translating frequently used APIs and aims to build a generic pipeline. Supporting all the libraries in the pipeline remains an open challenge \rebuttal{that we aim to address in future work}. 
Furthermore, while the idea of compositional translation and validation is PL-agnostic, the static analysis makes the extension of \approach to translating from other source projects challenging. Devising LLM-enabled or PL-agnostic static analysis approaches can benefit code translation approaches such as \approach. 
We also showed that the quality of the source project test suite can significantly impact the translation validation results. In future work, we plan to integrate an LLM-based test generator into the \approach pipeline to enhance the validation component.

\section{Data Availability}
\label{sec:data-availability}
Artifacts and implementation of \approach are publicly available~\cite{website,datarepository}. 

\begin{acks}
This work is supported by the IBM-Illinois Discovery Accelerator Institute and NSF CCF-2238045 grants. We thank Prof. Darko Marinov and Raju Pavuluri for their help with this research. We also thank the anonymous reviewers for their comments, which helped make this work stronger.
\end{acks}

\bibliographystyle{ACM-Reference-Format}
\bibliography{refs}

\end{document}